\documentclass{aastex631}

\shorttitle{Ap star catalog }
\shortauthors{Fangfei Shi et al.}
\graphicspath{{./}{figures/}}

\usepackage{amssymb} 
\usepackage{amsmath}  
\usepackage{times} 
\usepackage{color}

\usepackage{hyperref}
\hypersetup{hidelinks,
	colorlinks=true,
	allcolors=black,
	pdfstartview=Fit,
	breaklinks=true}
\begin{document}

\title{An Ap star catalog based on LAMOST DR9}

\author{Fangfei Shi}
\altaffiliation{fangfei1420@pku.edu.cn}
\affiliation{Department of Astronomy, School of Physics, Peking University, Beijing 100871, P. R. China}
\affiliation{Kavli Institute for Astronomy and Astrophysics, Peking University, Beijing 100871, P. R. China}

\author{Huawei Zhang}
\altaffiliation{zhanghw@pku.edu.cn}
\affiliation{Department of Astronomy, School of Physics, Peking University, Beijing 100871, P. R. China}
\affiliation{Kavli Institute for Astronomy and Astrophysics, Peking University, Beijing 100871, P. R. China}

\author{Jianning Fu}
\affiliation{Department of Astronomy, Beijing Normal University, Beijing 100875, P. R. China}
\affiliation{Institute for Frontiers in Astronomy and Astrophysics, Beijing Normal University, Beijing 102206, P. R. China}

\author{Donald Kurtz}
\affiliation{Centre for Space Research, Physics Department, North West University, Mahikeng 2745, South Africa}
\affiliation{Jeremiah Horrocks Institute, University of Central Lancashire, Preston PR1 2HE, UK}

\author{Maosheng Xiang}
\affiliation{Key Laboratory of Optical Astronomy, National Astronomical Observatories, Chinese Academy of Sciences, Beijing 100101, China}
\affiliation{Institute for Frontiers in Astronomy and Astrophysics, Beijing Normal University,  Beijing 102206,  China}

\begin{abstract} 
We present a sample of 2700 Ap stars in LAMOST DR9. The candidates are first selected to be in a temperature range typical of Ap stars by using the $BP$-$RP$ color index from {\it Gaia} DR3. Then the 5200\,\AA\ flux depression features characteristic of Ap stars are visually checked in LAMOST DR9 spectra. The detailed spectral features are given by applying a modified spectral classification program, MKCLASS. Stellar parameters of these Ap stars such as $T_{\rm eff}$, $\log g$, [Fe/H], [Si/H], and $v{\sin}i$ are either extracted from a hot star catalog or derived through empirical relations and then a statistical analysis is carried out. The evolutionary stages are also discussed. Finally, we discuss the rotation and pulsation features of those who have TESS or {\it Kepler} light curves. Among these Ap stars we find 7 new rotation variables, 1 new roAp star, and new $\delta$ Scuti pulsation of a previously known roAp star.
\end{abstract} 


\section{Introduction}
\label{intro}

Chemically peculiar (CP) stars are the main sequence stars that have peculiar surface abundances characterised by enhanced (or weakened) absorption lines \citep{1974ARA&A..12..257P}. Their spectral types cover from early-B to early-F, and depending on different temperatures and other atmosphere parameters, CP stars show a wide range of anomalies. It is generally accepted that some factors, such as magnetic field or binarity (e.g. \citealt{1974ApJ...191..165S,1973ApJS...25..137A,2016MNRAS.460.1912G}), make CP stars rotate slowly, reducing turbulent mixing so that atomic diffusion can efficiently work \citep{1970ApJ...160..641M}. However, these factors and how they govern the modification of the atmospheric abundances are not fully understood.

Many types of CP stars have been found. \citet{1974ARA&A..12..257P} firstly divided CP stars into four main subgroups and the characteristics were refined by several works (e.g. \citealt{1996A&A...310..872A}, \citealt{2018A&A...619A..98H}). 
The first subgroup is CP1 stars, always with weak Ca and/or Sc together with overabundances of iron-group elements, so they are also called metallic-line stars or AmFm stars. CP2 stars are the second subgroup. Most of them show some combination of enhanced Si, Cr, Sr, Eu, and rare earth elements, and sometimes, similar enhanced metal lines like CP1 stars. CP2 stars host large field and strong magnetic fields, which is believed to cause the peculiarities, so they are also called the magnetic Bp, Ap, and Fp stars (or generically Ap stars hereafter). The third subgroup is CP3 stars, also known as HgMn stars or Mercury-Manganese stars, since they always show enhanced Hg and/or Mn. CP4 stars are the last subgroup with weak He lines so that they are also called He-weak stars. Besides these four groups, $\lambda$ Bootis stars (underabundances of refractory elements; \citealt{2017MNRAS.466..546M}), Barium stars with abnormally strong lines of ionised barium \citep{1951ApJ...114..473B}, and the carbon stars \citep{1956VA......2.1428B} have also been discovered.

The Ap stars are the focus of the present research. The prime factor governing the peculiarities of Ap stars is the strong magnetic fields they host, typically roughly dipolar with field strengths of a kG, or more. This strong magnetic field suppresses convection, thus providing a stable environment for ions with many absorption features to be lifted by radiation pressure to the surface against gravity. However, some other elements, especially helium, sink deeper into the sea of hydrogen. 

Several catalogs of Ap and related stars have been published. After the first catalog of 6684 Ap and Am stars \citep{1991A&AS...89..429R}, \citet{2009A&A...498..961R} collected 8205 known or suspected Ap, HgMn, and Am stars from a large amount of literature. Large spectroscopic surveys provide more opportunities for us to study the spectral features of Ap stars. \citet{2019AandA...628A..81S} compiled a catalog containing 238 Ap stars through detecting magnetically split lines. In their catalog, 84 Ap stars are found by \citet{2017A&A...601A..14M} and 154 Ap stars are provided by \citet{2019ApJ...873L...5C} in Apache Point Observatory Galactic Evolution Experiment \citep[APOGEE, ][]{2017AJ....154...94M}. \citet{2019ApJS..242...13Q} compiled a catalog containing 9372 Am stars and flagged 1131 Ap candidates in the Large Sky Area Multi-Object Fiber Spectroscopic Telescope \citep[LAMOST, ][]{2012RAA....12..723Z,2012RAA....12.1197C} DR5, with the help of machine learning by identifying several enhanced lines. \citet[][hereafter HPB2020]{2020A&A...640A..40H} built a sample consisting of 1002 mCP (magnetic CP: most Ap and He-peculiar) stars by identifying the flux depression at 5200\,\AA\ and gave the spectral classification in the modified MKCLASS program. 

In this work, we introduce the data and method we used to build the catalog in Sections~\ref{data} and \ref{cat}, respectively, and then compare our sample with other literature in Section~\ref{compare}. The stellar parameters are derived in Section~\ref{param} and finally we discuss the photometric features in Section~\ref{pho}.

\section{Data}
\label{data}
The Large Sky Area Multi-Object Fiber Spectroscopic Telescope (LAMOST) is an optical spectral telescope located in Xinglong Station of the National Astronomical Observatory, China \citep{2012RAA....12..723Z}. 
The telescope is designed with 20 square degrees field of view and up to 4.9 meters effective aperture so that it is able to take spectra for 4000 targets with magnitude brighter than $r=19$ at a resolution $R \simeq 1800$ at one time in low resolution mode \citep{2012RAA....12..723Z}.
Taking advantage of the instrument, the LAMOST survey was launched in 2011 aiming to collect spectra for all the objects of interest covering from -10$^\circ$ to +90$^\circ$ declination. 
This is where our spectra come from. Up to DR9, LAMOST has obtained 11,791,589 spectra with spectral resolution $R \simeq 1800$ covering approximately a wavelength range from 3700 to 9000\,$\rm \AA$. 

 {\it Gaia} \citep{2016A&A...595A...1G} is a space telescope launched in 2013. It observes stars in a wide $G$ band ($330-1050$\,nm) together with a blue band, $BP$ ($330-680$\,nm) and a red band, $RP$ ($640-1050$\,nm) down to $G\sim 20.7$\,mag and also collects middle resolution spectra for the stars brighter than 16.2\,mag in $G$ band. With the astrometric and photometric measurements including distance, proper motion and magnitude of billions of the stars in the Milky Way, it can be used to study not only the stellar physics but also the formation and evolution of the Galaxy. Up to {\it Gaia} DR3, the astrometric and photometric measurements for nearly 2 billion sources brighter than $G = 21$ have been released \citep{2022arXiv220605989B}. In this work, the parallax and the color index, $BP-RP$, are obtained from {\it Gaia} DR3. 
 
 {\it Gaia} also gives stellar astrophysical parameters in DR3 \citep{2022arXiv220605541R,2022arXiv220606138A}. These parameters are measured in two ways: one is based on astrometry, photometry, and low-resolution spectra; the other is  based solely on spectra from the Radial Velocity Spectrometer (RVS) which gives middle resolution spectra ($R \simeq 11,500$). These parameters are used as input to produce further results, such as mass, age, and evolutionary stage.

In order to construct a large sample of Ap star candidates, a dataset of stellar spectra with high signal-to-noise ratios in the Sloan $g$ band (S/N$g \ge 50$) was chosen from LAMOST DR9, which includes 2,775,994 spectra. As the temperature range of Ap stars is above 6000\,K, we first selected stars with B, A, and F spectral classifications given by LAMOST. Since some stars do not have accurate LAMOST stellar classifications, we also considered the color index, $BP-RP$, given by {\it Gaia} DR3. According to Eric Mamajek\footnote{\href{https://www.pas.rochester.edu/~emamajek/EEM\_dwarf\_UBVIJHK\_colors\_Teff.txt}{https://www.pas.rochester.edu/{$\sim$}emamajek/EEM\_dwarf\_UBVIJHK\_colors\_Teff.txt} }, $T_{\rm eff}>  6000$\,K corresponds to an extinction corrected color, $(BP-RP)_0 < 0.767$. Here, the extinction was calculated through a new three-dimensional map of dust reddening, {\sc Bayestar} \citep{2019ApJ...887...93G}, based on galactic longitude $l$, galactic latitude $b$, and parallax from {\it Gaia} DR3. 
We select 929,043 spectra through $(BP-RP)_0$. Some stars do not have distance measurements, so it is impossible to calculate the de-reddened $(BP-RP)_0$. For these stars, we ignore the reddening. Their $(BP-RP)_0$ must be smaller after considering extinction thus $T_{\rm eff}$ must be higher than 6000\,K. So we also selected 2,926 spectra without distance measurements but with $(BP-RP) < 0.767$. In total, 931,969 spectra pass the color criterion.

\section{catalog}
\label{cat}
\subsection{Depression}
\label{depression}

There is a flux depression at 5200\,\AA\ that is essentially only seen in the spectra of magnetic CP stars  \citep{1998A&AS..128..573M,2018CoSka..48..218M}. This depression feature was firstly discovered by \citet{1969ApJ...157L..59K} together with the similar flux depressions at 4100 and 6300\,\AA\ through spectrophotometric observations of the Ap star HD\,221568.  
The combined contribution of Si, Cr, and Fe absorption at 5200\,\AA\ is the main cause of the depression \citep{2007A&A...469.1083K}. \citet{2014A&A...562A..65S} found that the three-filter $\Delta a$ photometric system \citep{1976A&A....51..223M} can be used to detect the peculiarity related to the 5200\,\AA\ region. 

In this system, two intermediate band filters ($g_1$ at 5020\,\AA\ and $g_2$ at 5240\,\AA) are used along with the Str\"omgren $y$ filter. The flux at $g_2$ is compared with that at $g_1$ and $y$. The index is introduced as:

$$a = g_2 -(g_1 +y)/2$$

\noindent and it describes the depth of the flux depression at 5200\,\AA. As shown in \citet{2014A&A...562A..65S}, the index $a$ depends on color index, in other words, depends on $T_{\rm eff}$. The relationship between index $a$ and $(B-V)_0$ color is nearly linear for bluer stars. 
To compare $a$ values of peculiar stars and non-peculiar stars, an intrinsic peculiarity index, $\Delta a$, is used:
$$\Delta a=a-a_0,$$
\noindent where $a_0$ is the index for non-peculiar stars with the same color. Then the peculiar stars are selected with a large enough $\Delta a$. This is a traditional way to examine this depression feature, as HPB2020 used in their work: they calculated $\Delta a$ by comparing $a$ indices with those of normal stars and selected the stars with $\Delta a$ larger than +75\,mmag as their mCP star sample. 

\begin{figure}
\centering
\includegraphics[width=0.44\linewidth,angle=0]{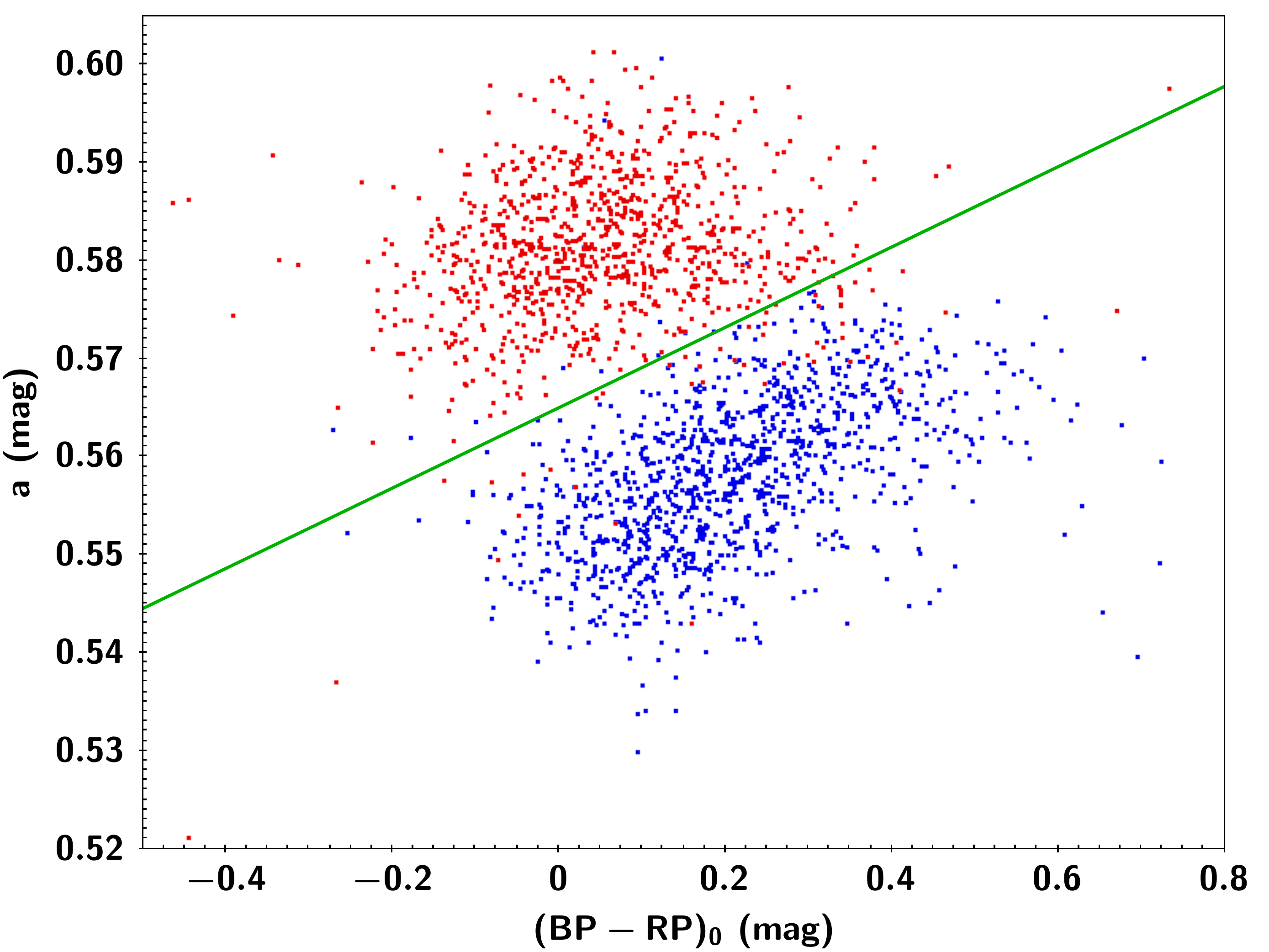}
\includegraphics[width=0.44\linewidth,angle=0]{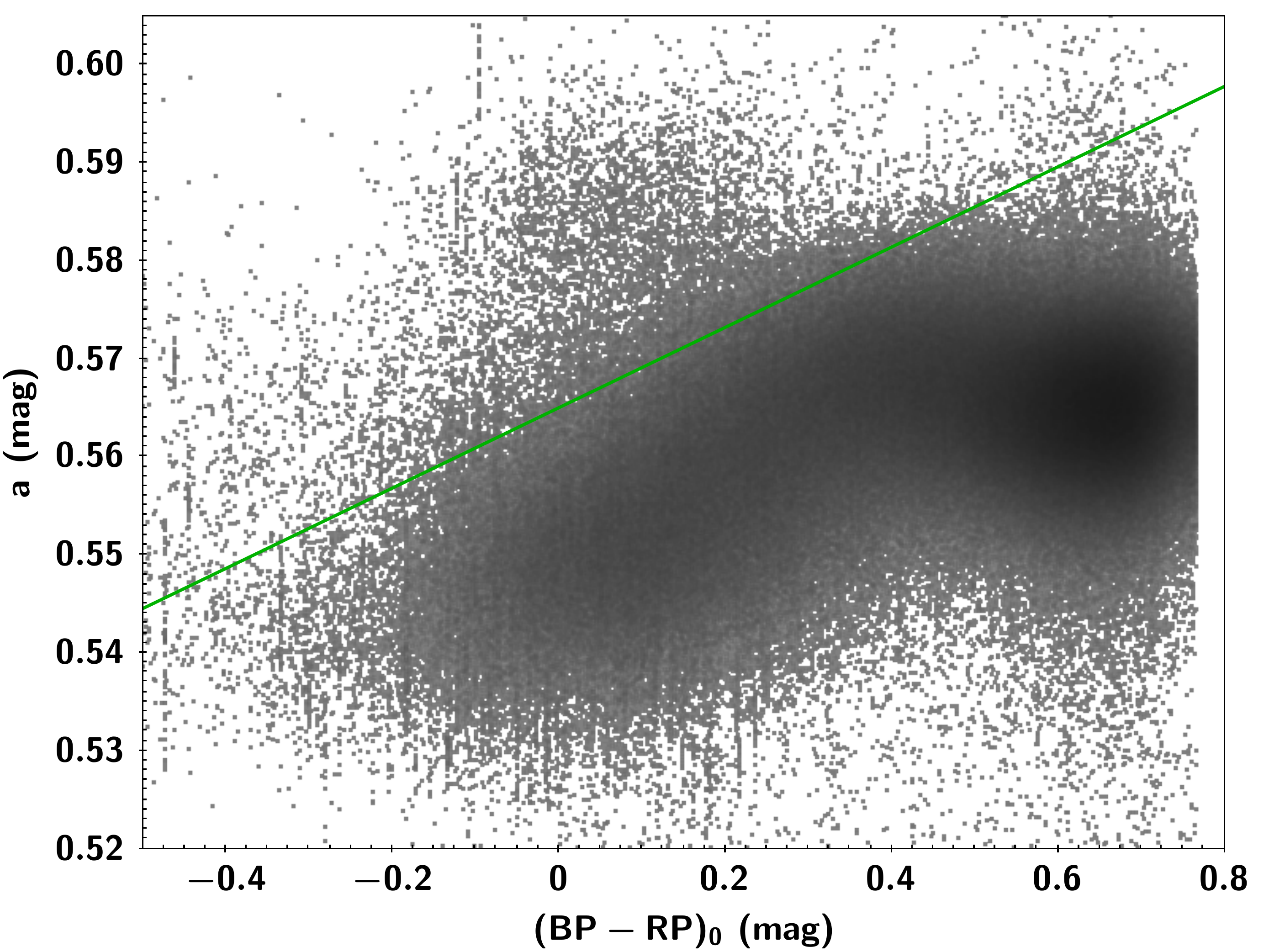}		
\caption{Left: The distribution of Ap stars (red) and non-peculiar stars (blue) on the $(BP-RP)_0$ versus $a$ diagram. The green line separates the two groups. 
Right: The distribution of the spectra selected by $(BP-RP)_0$ (or $(BP-RP)$) and $T_{\rm eff}$ in Section~\ref{data} on the $(BP-RP)_0$ (or $(BP-RP)$) versus $a$ diagram. The green line is the same as that above. }
\label{fig:mcp_line}
\end{figure}

 
In Fig.~\ref{fig:mcp_line} we plot Ap stars and normal stars in the ($(BP-RP)_0$, $a$) diagram with the expectation that a line can separate them, since Ap stars should have larger $a$ compared to non-peculiar stars with the same color, $(BP-RP)_0$. To test this, we selected mCP stars in HPB2020 and chose 1000 normal stars in LAMOST DR4 that are not in the HPB2020 samples. 
$(BP-RP)_0$ and $a$ were calculated for these stars. To calculate the magnitudes in each filter, the spectra were normalised to the flux at 4030\,\AA\  and then folded with the filter curves defined in \citet{2003MNRAS.341..849K}. The left panel in Fig.~\ref{fig:mcp_line} shows that we do find a line that separates the two groups.

There are 931,969 spectra that pass the color criterion are also plotted in the $(BP-RP)_0$ (or $(BP-RP)$) versus $a$ diagram (right panel in Fig.~\ref{fig:mcp_line}). In the diagram, there are 10,750 spectra above the line. This indicates that these spectra have significant deep absorption around 5200\,\AA\ .
  
\begin{figure}
\centering
\includegraphics[width=0.55\linewidth,angle=0]{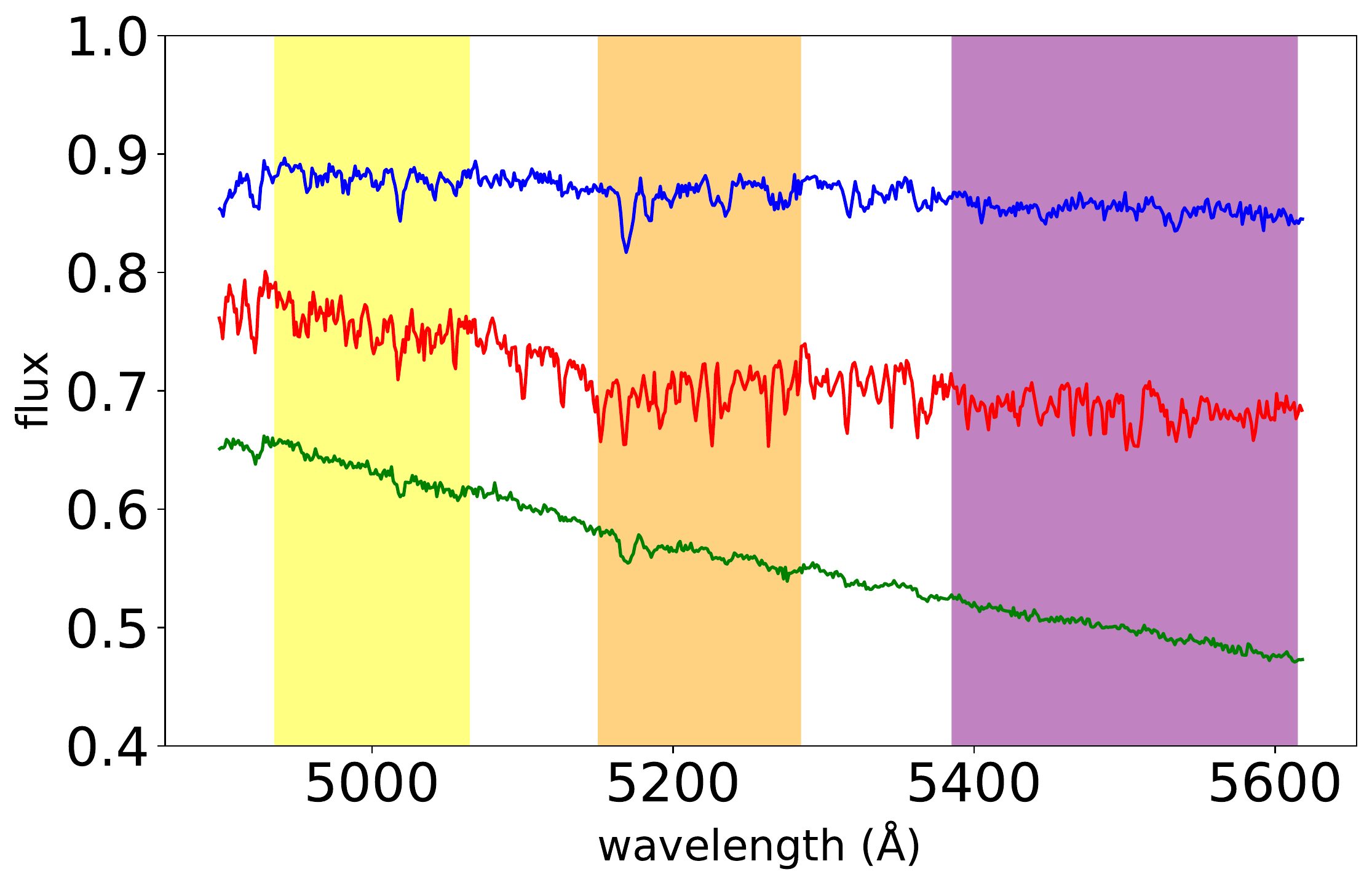}	
\caption{The 4900 to 5600\,\AA\  region of example spectra of an Ap star (red spectrum in middle) and a normal star (green spectrum in bottom) with similar $T_{\rm eff}$. The blue one (top), as an example, shows the spectrum of a star whose large $a$ index is caused by absorption lines in $g_2$ band rather than a blended absorption thus it is removed after inspection. The main ranges of $g_1$, $g_2$, and $y$ filter are shadowed in yellow, orange, and purple, respectively. The spectra are normalized at 4030\,$\rm \AA$. For clarity, the continuum of the Ap stars has been offset by 0.1.}
\label{fig:Ap_example}
\end{figure}

Since the depression has the multiple contributions from Si, Cr, and Fe, it is more likely to be a blended absorption (red one in Fig.~\ref{fig:Ap_example}) rather than lines (blue one in Fig.~\ref{fig:Ap_example}). For insurance, all the 10,750 spectra passing the index $a$ criterion are inspected one by one to make sure that the selected spectra have obvious depression features and enhanced absorptions at several Si, Cr, Sr, and Eu lines. For one star, only the spectrum with the highest signal-to-noise ratio was chosen. After this, we have 2700 individual Ap stars remaining.

However, the 5200\,\AA\ depression is only an indicator of enhanced Si, Cr or Fe, and this feature in some Ap stars may not be significant enough to be selected by our criteria, so our Ap stars are biased towards those with significant flux depressions at 5200\,\AA. Some Ap stars with ambiguous depression features may also be missed in this process since we want to make sure of a pure Ap star sample.

\subsection{Peculiarity features}
\label{pec}

MKCLASS is an automatic classification code designed to classify spectra following the same way as the traditional classification process done by humans \citep{2014AJ....147...80G}. The current MKCLASS (v1.07) was developed to identify spectral features in the range of $3800-5600$\,\AA\ by comparing the observed spectra with standard ones in a specific library. The spectra in different libraries were obtained with different resolutions and wavelength ranges. MKCLASS is able to give not only the spectral and luminosity classification but also a few sets of spectral peculiarities, so that it can help to search for CP stars. 
For example, \citet{2016AJ....151...13G} applied the \mbox{MKCLASS} to classify the spectra in the LAMOST-{\it Kepler} project and found 1067 Am stars in 3088 A4 to F1 dwarfs and subgiants. Currently, the lines related with Ap stars which MKCLASS can identify are the blend contributions from Si II, Sr II, and Cr II at 4077\,\AA, the blend of Si II and Eu II at 4130\,\AA, and the Eu II 4205\,\AA\ line.

Considering the special features of Ap stars, we modified the MKCLASS code following the methods developed by HPB2020. Apart that we abandon two Si\,II lines beyond 5600{\,\AA} and He\,I lines, we use the same line list including two blend lines (4077\,\AA, 4130\,\AA), four Si\,II lines (3856\,\AA, 4200\,\AA, 5041\,\AA, 5056\,\AA), two Cr\,II lines (3866\,\AA, 4172\,\AA), one Sr\,II line (4216\,\AA), and one\,Eu II line (4205\,\AA). 
The original library, libnor36, and the library built by HPB2020 with LAMOST spectra, liblamost, are used because both have similar resolution and flux processing as the spectra in LAMOST.
The detailed spectral classifications are given in the catalog in the form of element+wavelength following the work of HPB2020.

The two libraries do not always assign for the same star the same spectral type.
The typical uncertainty of the spectral type is $\pm1$\,subclass, and for some stars, this increases up to $\pm2$\,subclasses.
Considering that the peculiarities in Ap star spectra make the classification difficult, we use 2 subclasses as a criterion for the reliability of spectral classifications. If the deviation of the 2 spectral classifications is not larger than 2, then the classification results are considered reliable. For those that are less reliable the classification results are flagged with star marks in Column 8-9 in the catalog.


Table~\ref{Tab:cat} shows an extract of our catalog of the total 2700 Ap stars and is organised as follows: 

\begin{itemize}
\item[-] Column 1: observation ID in LAMOST,
\item[-] Column 2-3: RA and DEC,
\item[-] Column 4: absolute $G$ magnitude estimated from {\it Gaia} DR3 apparent $G$ magnitude and distance,
\item[-] Column 5: de-reddened $(BP-RP)_0$ (or $(BP-RP)$), 
\item[-] Column 6: $a$ index,
\item[-] Column 7: distance given by \citet{2021AJ....161..147B},
\item[-] Column 8-9: spectral classification using the standard star libraries liblamost and libnor36, respectively,
\item[-] Column 10: rotation and pulsation features,
\item[-] Column 11: {\it Gaia} id.
\end{itemize}

\section{Comparison with other catalogs}
\label{compare}

There are other peculiar star catalogs in the literature. In this section, our samples are compared with these catalogs and the numbers of the stars in common in each catalog are listed in Table~\ref{Tab:common}.

\begin{table*}

\scriptsize
\centering
\caption{The numbers of the stars in common in each catalog.} 
\setlength{\tabcolsep}{6mm}{
\begin{tabular}{lllll}
\hline
\hline
catalog & Numbers of stars & Numbers of stars & Numbers of stars & Numbers of stars \\
 & in total & within the color range & above the line in Figure~\ref{fig:mcp_line} & in our Ap catalog\\
\hline
Renson, Gerbaldi $\&$ Catalano (1991) & 6684 & 125 & 59 & 52 \\
Renson $\&$ Manfroid (2009) & 3652 & 134 & 70 & 68 \\
\citet{2019AandA...628A..81S} & 237 & 24 & 20 & 20 \\
\citet{2019ApJS..242...13Q} all & 9372 & 9158 & 573 & 65\\
\citet{2019ApJS..242...13Q} with Ap flag & 1131 & 995 & 130 & 44\\
HPB2020 & 1002 & 996 & 935 & 927\\

 \hline
 \hline
\end{tabular}}
\label{Tab:common}
\end{table*}

\citet{1991A&AS...89..429R} and \citet{2009A&A...498..961R} are two editions of catalogs including Ap stars. 
The former contains 6684 Ap and Am stars while the latter contains 3652 Ap stars.
 After crossmatching with our work, there are only 52 and 68 stars in common, respectively. Their work was mainly for bright stars which are not the targets of LAMOST, so there are only 270 stars observed by LAMOST. About half of these stars have large $a$ index and passed our depression inspection. The other half of the stars are missed. As discussed in Section~\ref{cat}, our method based on depression features can not perfectly select all the Ap stars.


\citet{2019AandA...628A..81S} built a catalog which contains 237 Ap stars by measuring magnetic fields or resolved magnetically split lines. Since the most obvious feature which can distinguish Ap stars from other CP stars is magnetic fields, their catalog provided a reliable Ap star samples. Twenty-four stars are also selected through color by us and among them 20 stars have large $a$ indice and selected as Ap stars in this work.

\citet{2019ApJS..242...13Q} published 9372 Am stars in LAMOST DR5, and flagged 1131 Ap star candidates. Sixty-five stars are in common with our stars, and among them 44 stars are flagged as Ap stars in their work. As discussed in HPB2020, \citet{2019ApJS..242...13Q} tended to find CP1 (Am) stars with the large differences ($\ge 5$) between the spectral subtypes obtained from Ca lines and H lines given by MKCLASS, which is not a typical feature of Ap stars.

We also compare our sample with HPB2020. Both works identified Ap stars using the 5200\,\AA\ depression, thus large overlap between these two samples can be expected. There are 927 stars in common with their 1002 mCP stars. 
Four stars are missed after crossmatching with {\it Gaia} DR3 using positions within 4\,arcsec; for insurance, we did not include them in our sample. 
Sixty-three stars do not have large enough $a$ index and eight stars do not pass our inspection process. As shown in the left panel of Figure~\ref{fig:mcp_line}, some Ap stars are located below the criteria line which means our catalog may miss some Ap stars with unconspicuous $a$ index.

There are 1147 stars in our candidates that are present in LAMOST DR4 but were not included in 1002 mCP stars from HPB2020. 
To make sure these stars have peculiarities, their spectra are visually checked by comparing with the spectra of the standard stars in libnor36 library with the similar spectral type (the spectral types are given by MKCLASS).
Among these 1147 stars, 16 stars are also in \citet{1991A&AS...89..429R} and \citet{2009A&A...498..961R}. It is therefore possible that HPB2020 may have missed more Ap stars. Another difference between our two methods is that when examining the {5200\,\AA } depression HPB2020 compared this absorption feature with some standard stars while we use an empirical line on the $(BP-RP)_0$ versus $a$ diagram. 
Our methods may let more stars pass to the visually inspection process, so that more Ap stars are found. 
Among our 2700 Ap stars, 2074 stars were observed before LAMOST DR4 which make up to 77$\%$ of the total. This is reasonable because more than 66$\%$ of the DR9 data was observed before DR4. 

\section{stellar parameters}
\label{param}

Some statistical properties of our final Ap stars are discussed in this section. The statistic distributions of several parameters, such as $T_{\rm eff}$, $\log g$, mass, and absolute magnitude are shown first. Then $v\sin i$, [Fe/H], and [Si/H] of Ap stars are compared with normal stars. Finally, the evolution stages are discussed.

\subsection{$T_{\rm eff}$, $\log g$, mass, and absolute magnitude }
\label{teff}

In this section, we show how to derive the stellar parameters for the Ap stars investigated in this paper and discuss the distribution of their parameters.
$T_{\rm eff}$ values were extracted from \citet{2021arXiv210802878X}, which provides the basic stellar parameters for 352,987 hot stars from LAMOST DR6. They developed a spectral fitting tool called {\sc The Payne} to derive 11 parameters including $T_{\rm eff}$. Using this hot star catalog, we extract $T_{\rm eff}$ values for 2550 stars in our Ap star catalog.

\begin{figure}
\centering
\includegraphics[width=0.45\linewidth,angle=0]{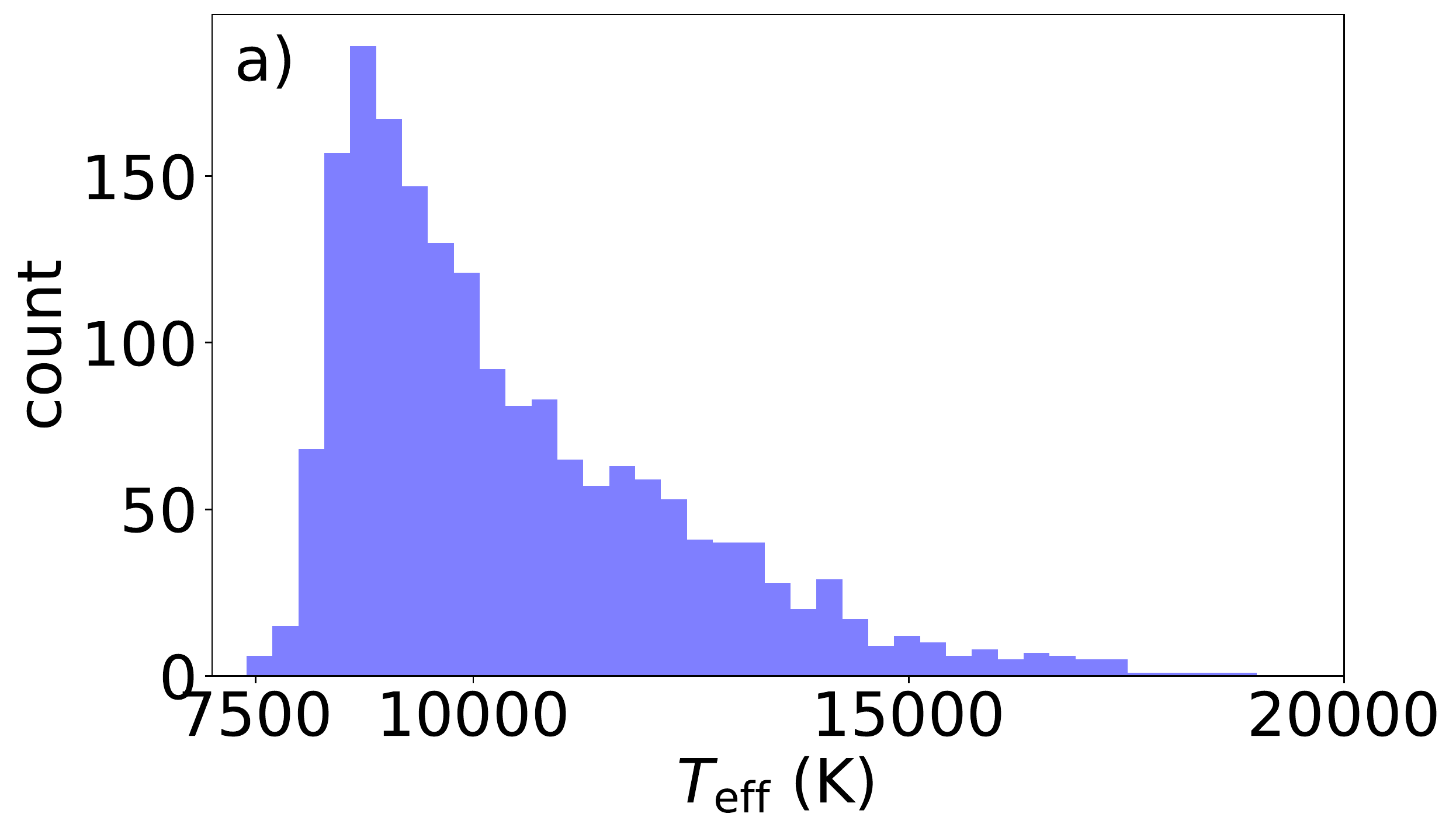}	
\includegraphics[width=0.45\linewidth,angle=0]{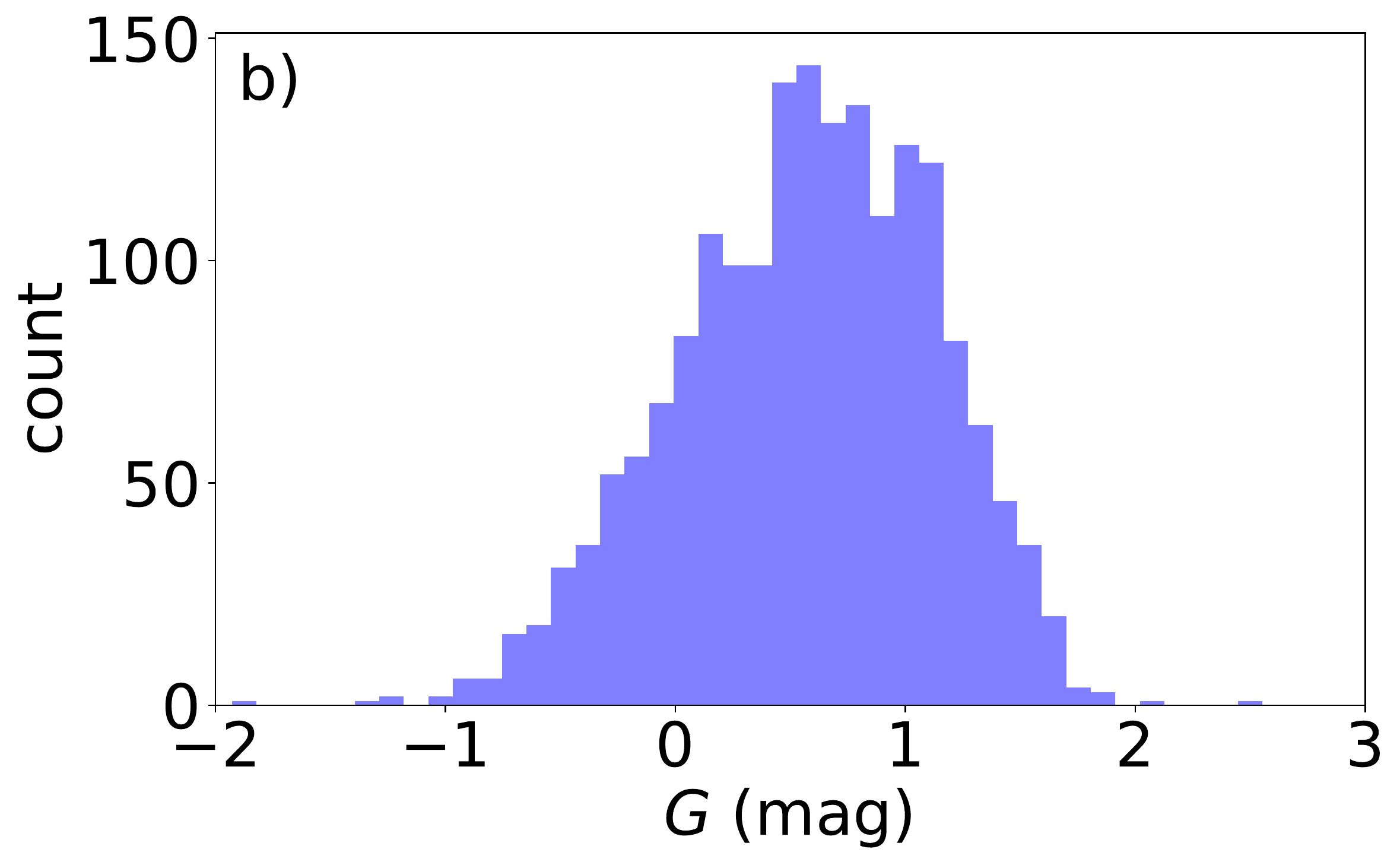}	
\includegraphics[width=0.45\linewidth,angle=0]{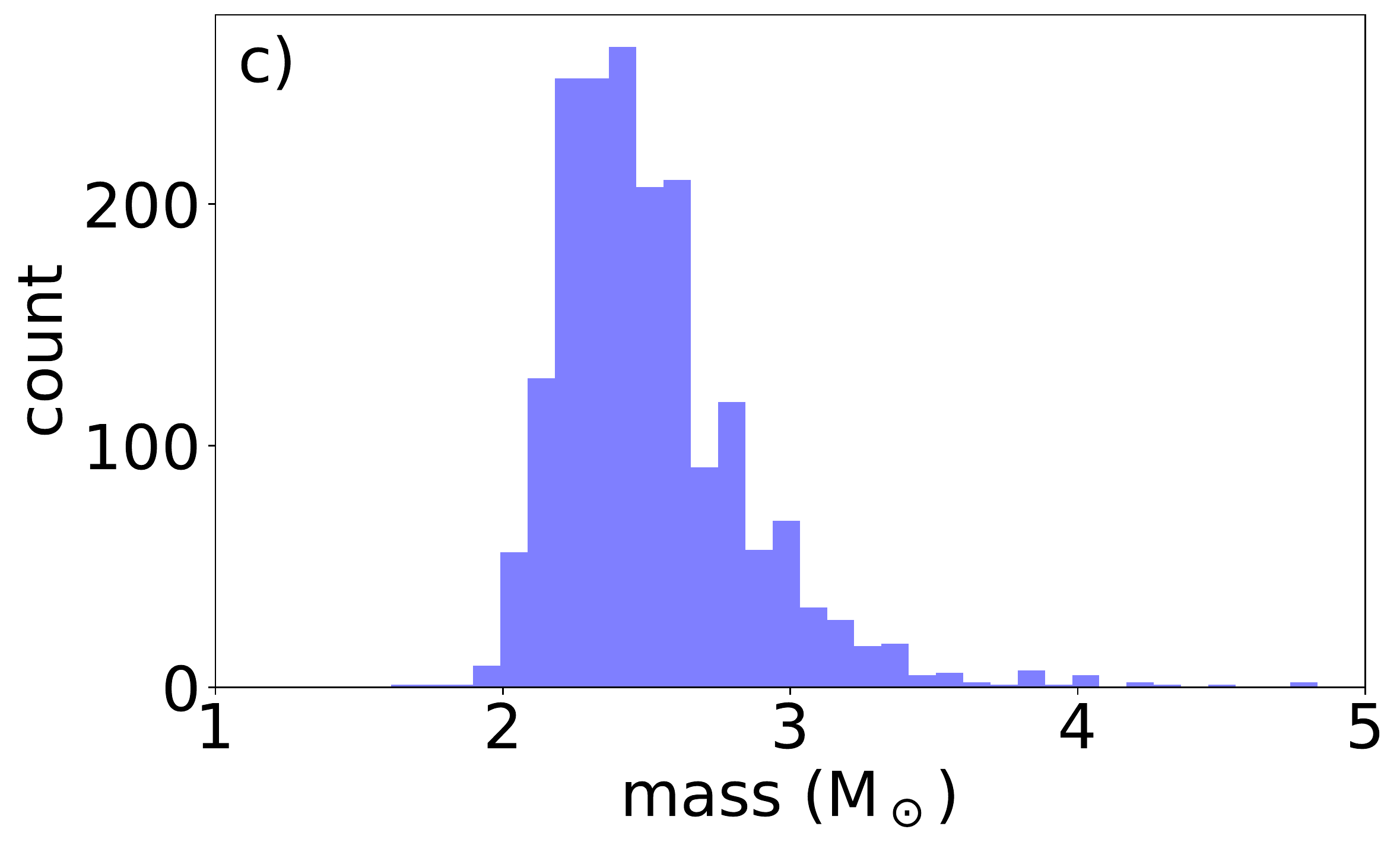}	
\includegraphics[width=0.45\linewidth,angle=0]{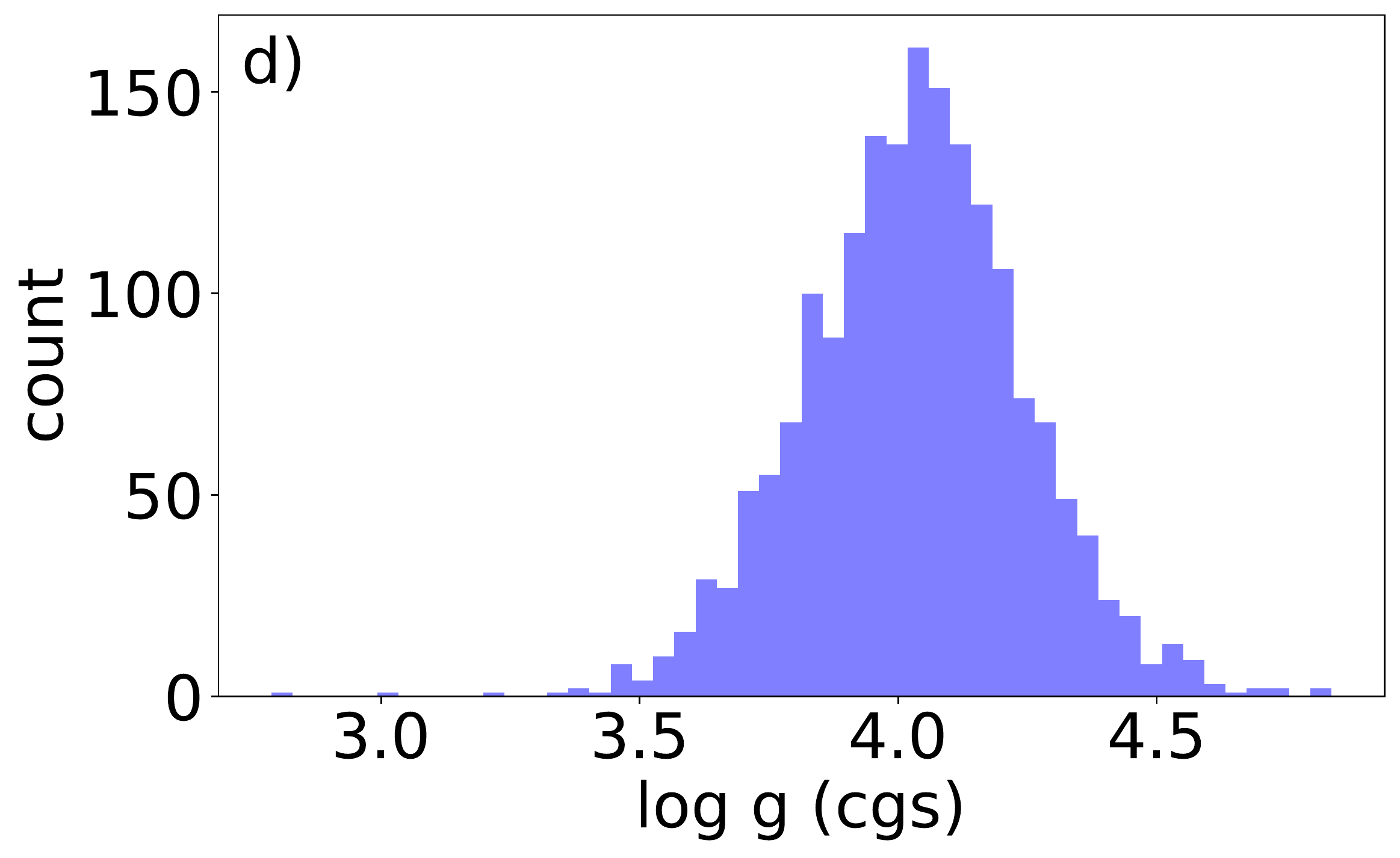}
\caption{Histograms of the effective temperatures (a), absolute $G$ magnitudes (b), masses (c), and $\log g$ (d) values on the main sequence of our Ap stars. }
\label{fig:param_hist}
\end{figure}

Panel a in Fig.~\ref{fig:param_hist} shows the histogram of $T_{\rm eff}$ of our Ap stars. The distribution has a peak around 8000\,K. None of these stars have $T_{\rm eff}$ lower than F stars ($T_{\rm eff}<7500\,$K) or higher than B stars ($T_{\rm eff}>30000\,$K). 

\begin{figure}
\centering
\includegraphics[width=0.6\linewidth,angle=0]{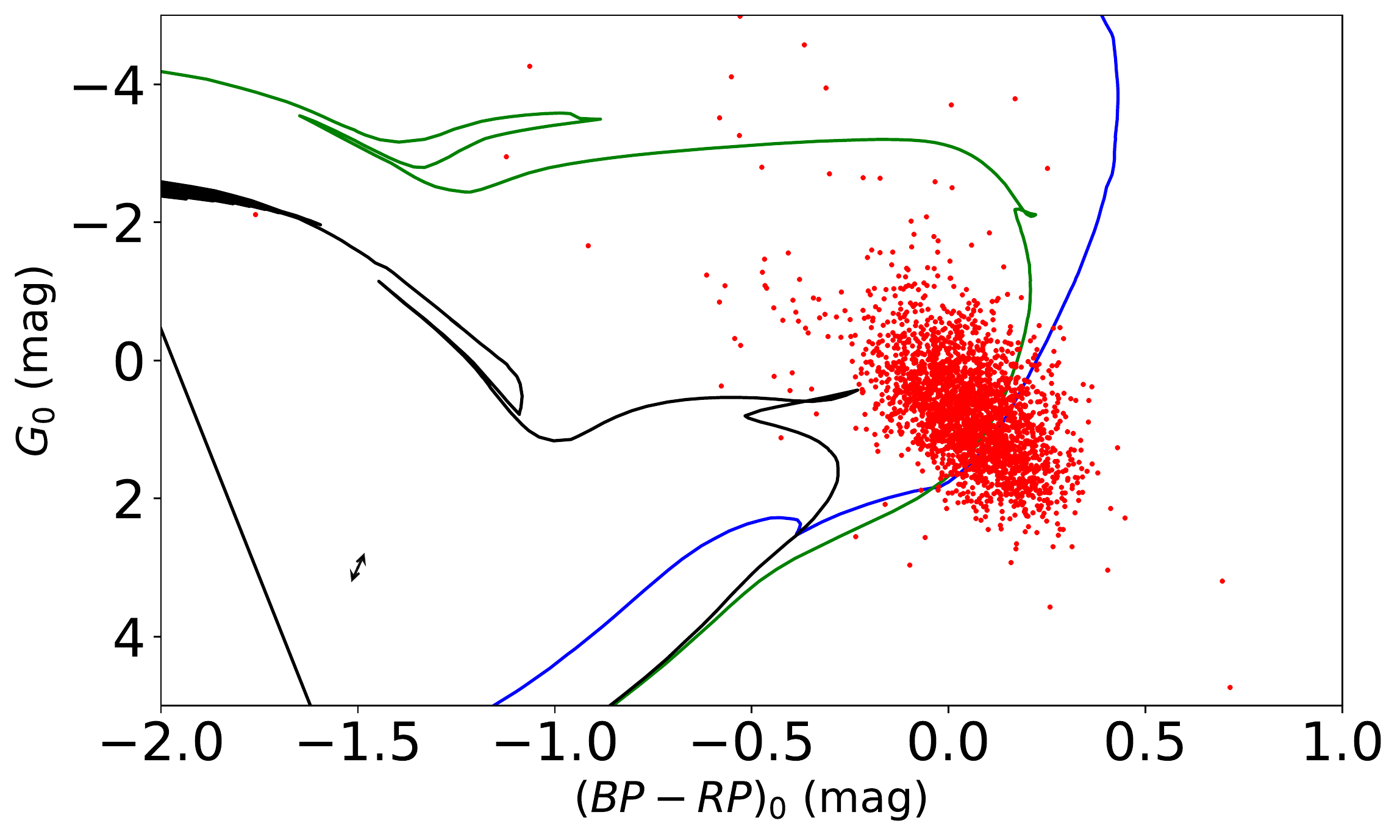}	
\caption{Ap stars distribution on $(BP-RP)_0$ versus $G_0$ diagram. The blue, green, and black lines are isochrones calculated with ages $10^7$, $10^8$, and $10^9$\,yr, respectively. The arrow shows the typical uncertainty of the position in this diagram.}
\label{fig:hr}
\end{figure}

To calculate the absolute $G$ magnitudes, extinctions were obtained through {\sc Bayestar}, as mentioned in Section~\ref{data}, and the distances were extracted from {\it Gaia} DR3 \citep{2021AJ....161..147B}. There are two types of distances given in their work, one is the geometric distance which is derived from parallaxes, and the other is the photo-geometric distance which additionally uses the color and apparent magnitude of a star. 
According to \citet{2021AJ....161..147B}, although two distance measurements give similar results at small distance, the photo-geometric distances perform better at larger distance. Therefore, the photo-geometric distance is used to calculate the extinction. Panel b in Fig.~\ref{fig:param_hist} shows the histogram of the absolute G magnitude of our Ap stars.

The $(BP-RP)_0$ versus $G$ diagram is presented in Fig.~\ref{fig:hr}. To estimate the typical uncertainty of the positions on this diagram, the median uncertainties of all parameters given by {\it Gaia} DR3 were taken as their typical uncertainties. The typical uncertainties of parallax, $G$, $BP$, and $RP$ are 0.02\,mas, 0.003\,mag, 0.003\,mag, and 0.004\,mag, respectively. 
The typical uncertainty of the extinction in $(BP-RP)_0$ is similar to that in $b-y$, which is about 0.1\,mag \citep{2019ApJ...887...93G}. The arrow in Fig.~\ref{fig:hr} indicates the typical uncertainty of the position on this diagram after considering reddening and parallax. 

The mass values were determined through fitting isochrones provided by PARSEC \citep{2012MNRAS.427..127B} on the $(BP-RP)_0$ versus $G_0$ diagram. Since most of the positions of stars overlap with both main sequence and pre-main sequence stages, and no evolved stars are found, we assume that these stars are all in the main sequence phase. 
Here we fit the mass values only for the stars whose positions have a distance of less than 0.05 mag to the theoretical grid point, since this is a ``good fit" distance used in HPB2020. Under this criterion, the mass values of 1841 stars are derived. 
For the metallicity [Z], we choose to use [Z] = 0.02 \citep{2019Atoms...7...41V} which has a good agreement with helioseismology rather than the latest solar abundance, [Z]=0.014 \citep{2021A&A...653A.141A}. 
This is reasonable because the Ap stars are supposed to have higher metallicities since they are younger.
The age range also follows HPB2020 and is set to be 1\,Myr to 10\,Gyr.
Panel c in Fig.~\ref{fig:param_hist} shows the histogram of the masses of our Ap stars. Most Ap stars have masses in the range of $2-3$\,M$_{\odot}$, which is the typical mass range of Ap stars. 

\begin{figure}
\centering
\includegraphics[width=0.6\linewidth,angle=0]{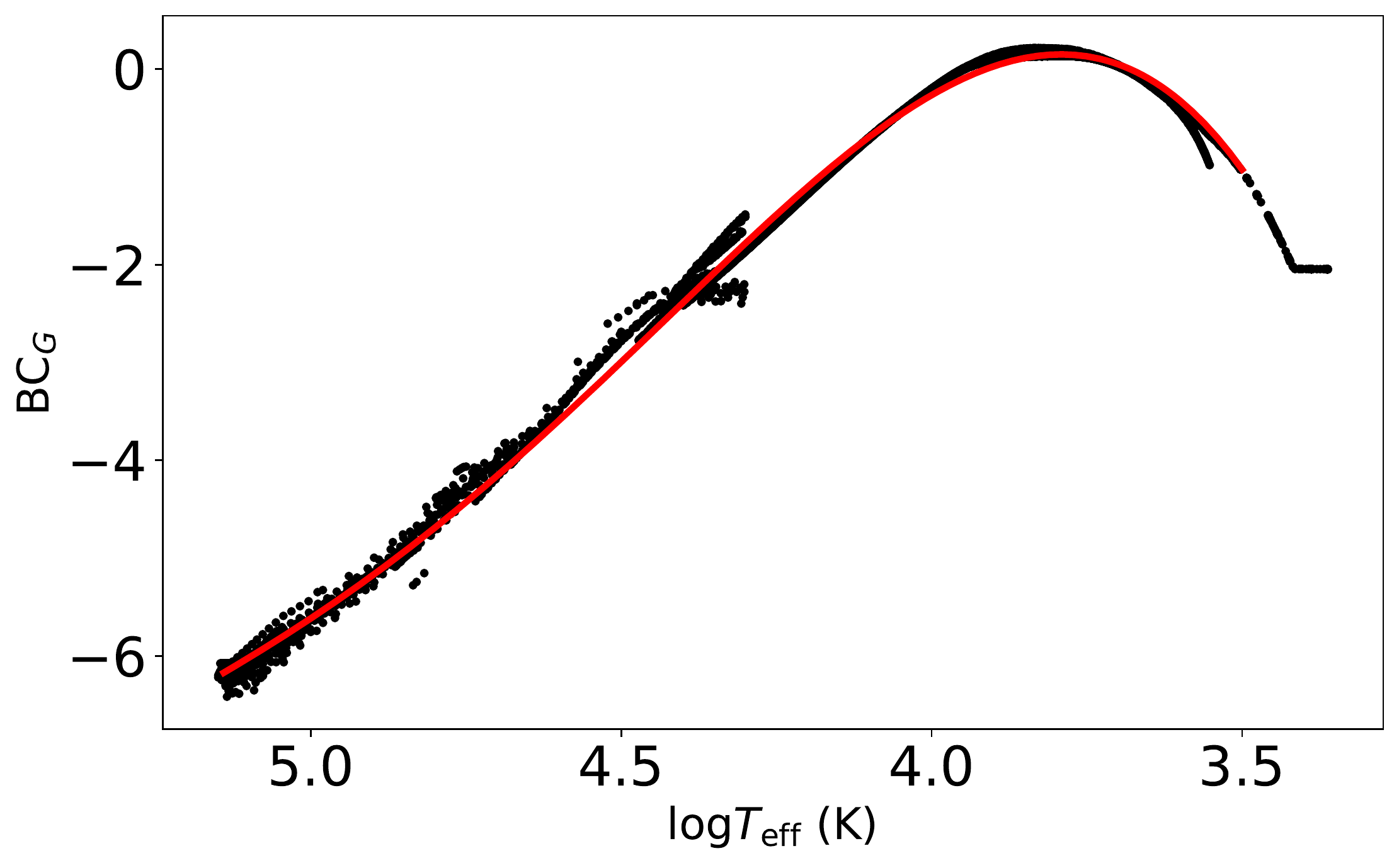}	
\caption{The relationship (red line) between $BC_G$ and $T_{\rm eff}$, which is the fit curve of the data given by isochrones (black dots).}
\label{fig:bc_teff}
\end{figure}

The $\log g$ values were calculated based on a mass -- luminosity relationship \citep{2005MNRAS.364..712Z} which comes from $L = 4\pi R^2 \sigma T^4$ and $g = GM/R^2$:
\begin{equation}
\log\frac{g}{{\rm g_{\odot}}} = \log \frac{M}{{\rm M_{\odot}}} + 4\log \frac{T_{\rm eff}}{{\rm T_{\rm eff,\odot}}} + 0.4(M_{\rm bol}-{\rm M_{bol,\odot}}),
\end{equation}

\noindent where $M_{\rm bol}=G + BC_G$. The $BC_G$ values were estimated from the effective temperatures. The isochrones give many sets of $BC_G$ and $T_{\rm eff}$ values, and with them the relationship between $BC_G$ and $T_{\rm eff}$  is plotted and fitted in Fig.~\ref{fig:bc_teff}. 
Since the spectral energy distribution of both hot stars and cool stars is more concentrate in infrared and ultraviolet range, respectively, the bolometric correction is large and negative, which is consistent with the relationship shown in Fig.~\ref{fig:bc_teff}. The $BC_G$ values of our Ap stars are given based on the fitted relationship and $T_{\rm eff}$ in \citet{2021arXiv210802878X}.

Panel d in Fig.~\ref{fig:param_hist} shows the histogram of the $\log g$ values of our Ap stars. The $\log g$ of these stars are in the range of main-sequence stars. 

\subsection{$v\sin i$, [Fe/H], [Si/H]}

\begin{figure}
\centering
\includegraphics[width=0.33\linewidth,angle=0]{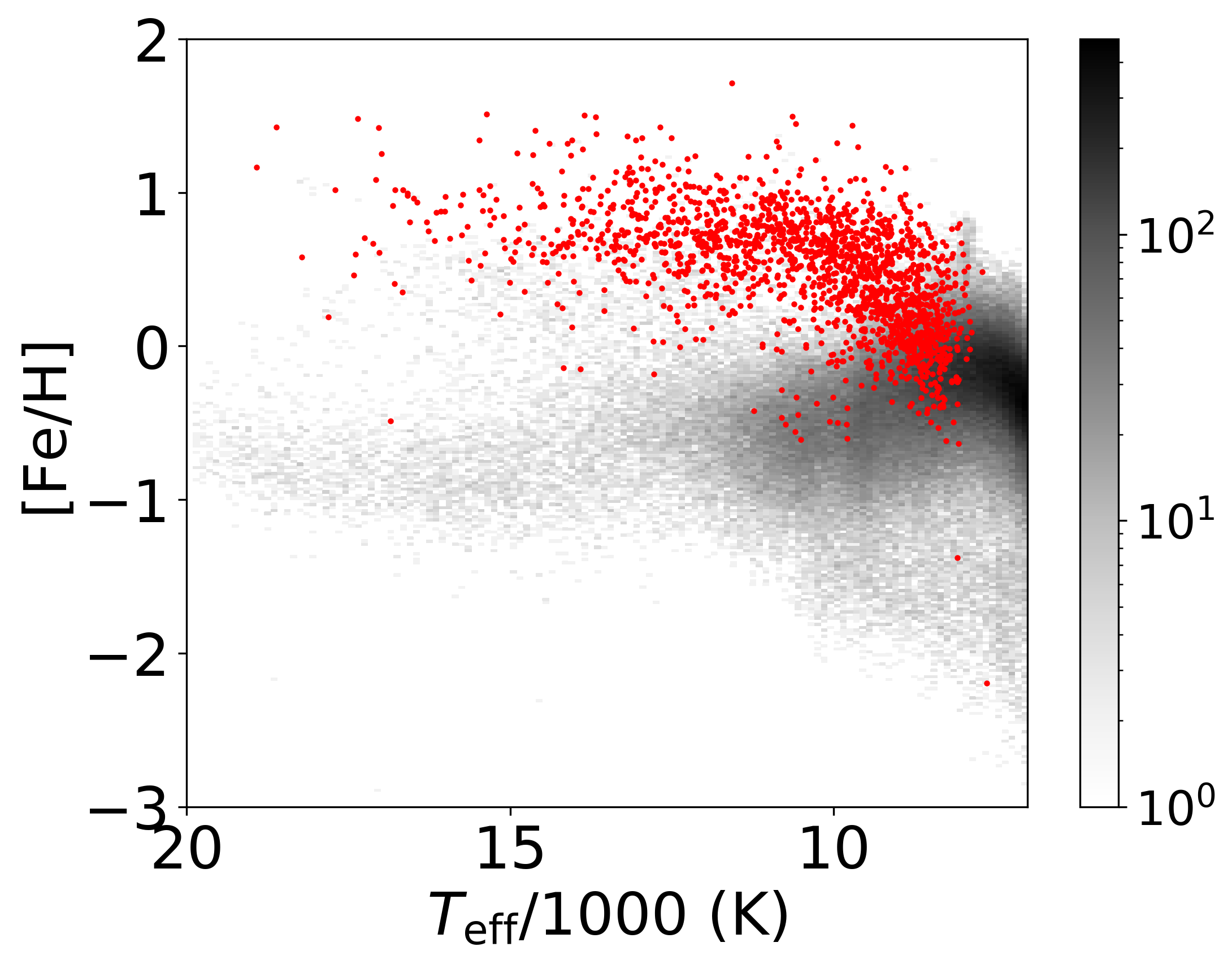}
\includegraphics[width=0.31\linewidth,angle=0]{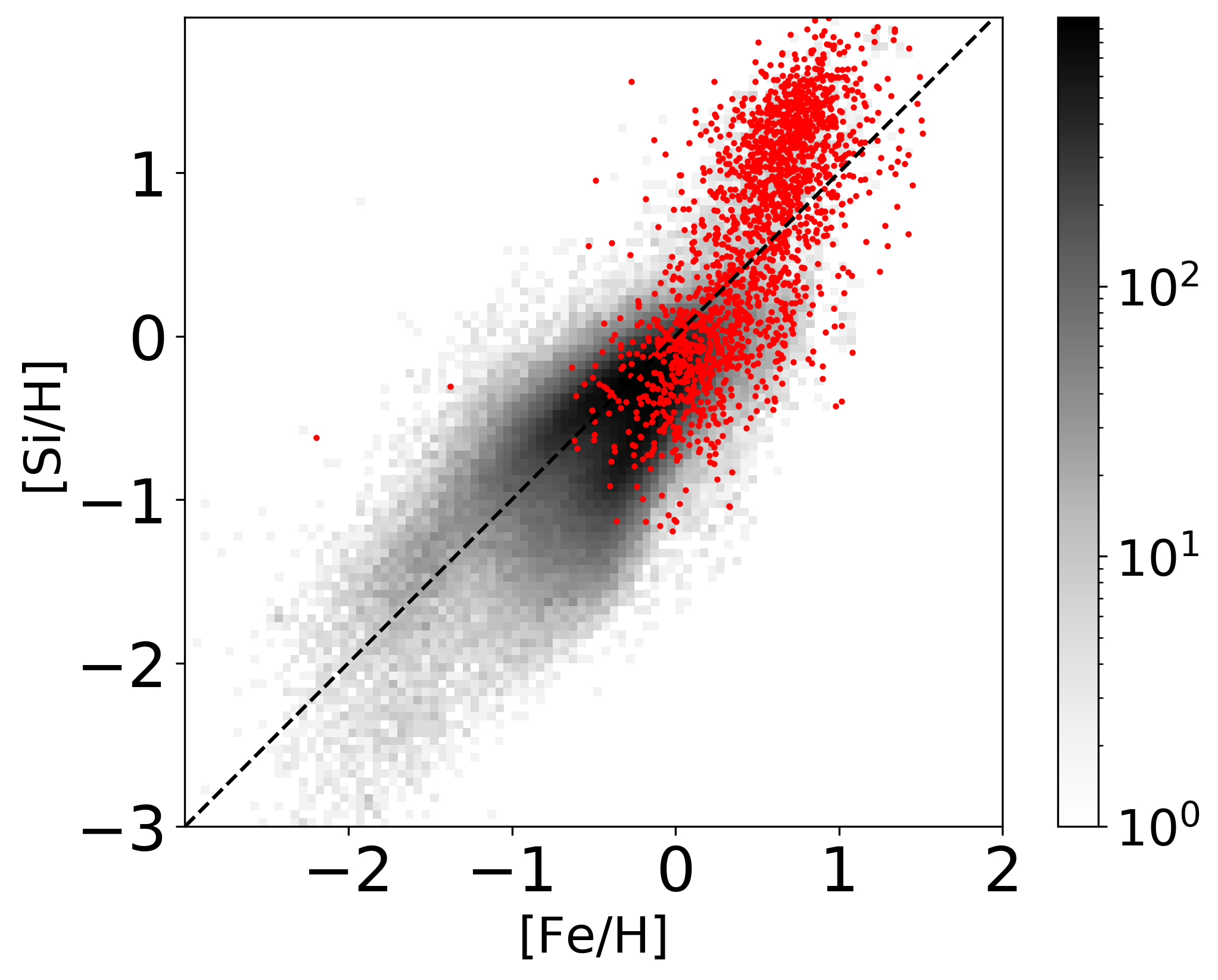}
\includegraphics[width=0.33\linewidth,angle=0]{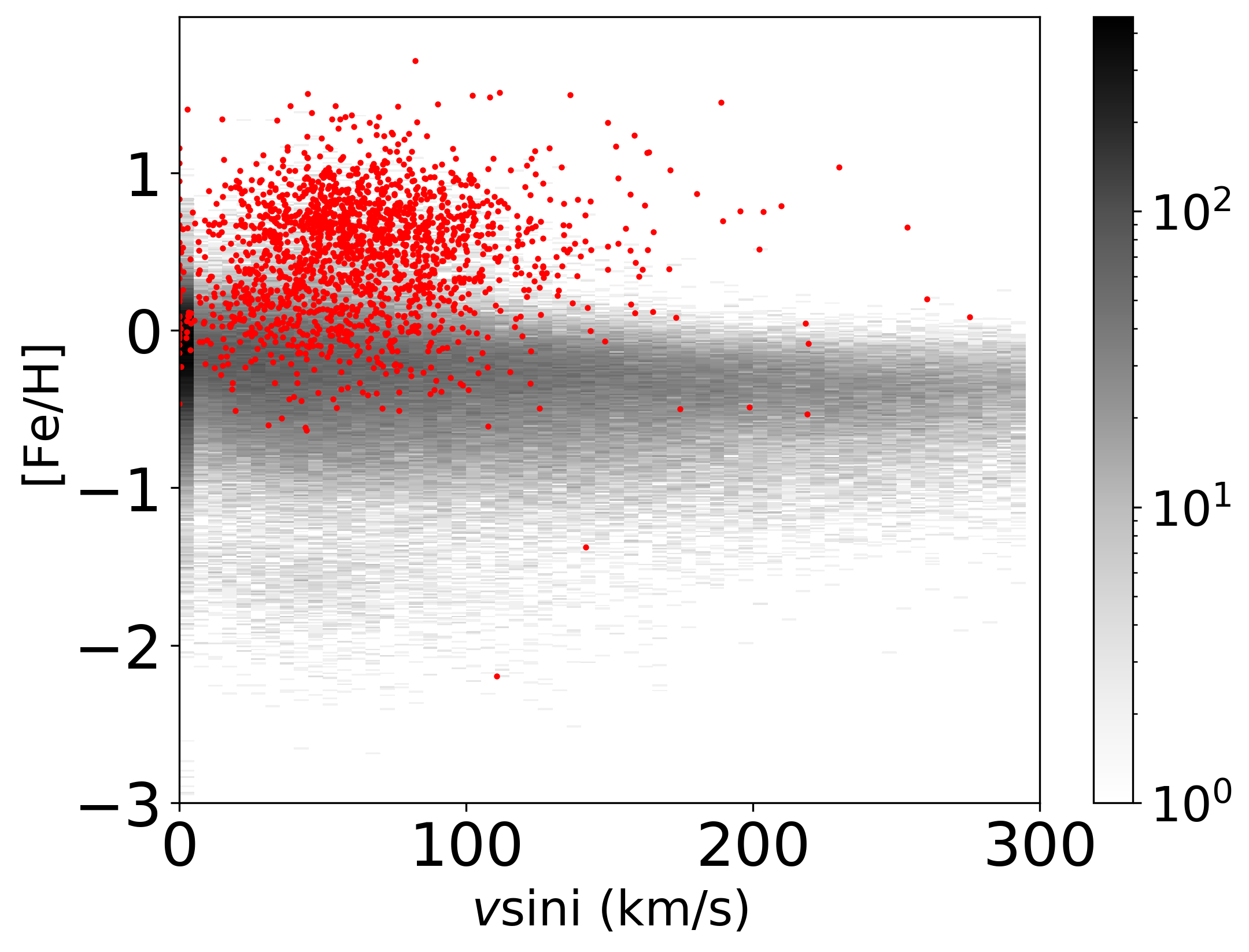}
\caption{The distribution of our Ap stars (red dots) and the density distribution of the stars in \citet{2021arXiv210802878X} (grey dots) in the $T_{\rm eff}$ -- [Fe/H] (left), [Si/H] -- [Fe/H] (middle), and $v\sin i$ -- [Fe/H] (right) planes.  }
\label{fig:plane}
\end{figure}

$T_{\rm eff}$, [Fe/H], [Si/H], and $v\sin i$ were extracted from \citet{2021arXiv210802878X}. [Fe/H] -- $T_{\rm eff}$, [Si/H] -- [Fe/H], and [Fe/H] -- $v\sin i$ of our Ap stars were plotted and compared with those about 452,000 stars in the catalog of \citet{2021arXiv210802878X}. In Fig.~\ref{fig:plane}, the majority of our Ap stars exhibit high [Fe/H], and show high [Si/H] and low $v\sin i$, which are typical features of Bp and Ap stars and consistent with the inference in \citet{2021arXiv210802878X}.

\begin{figure}
\centering
\includegraphics[width=0.32\linewidth,angle=0]{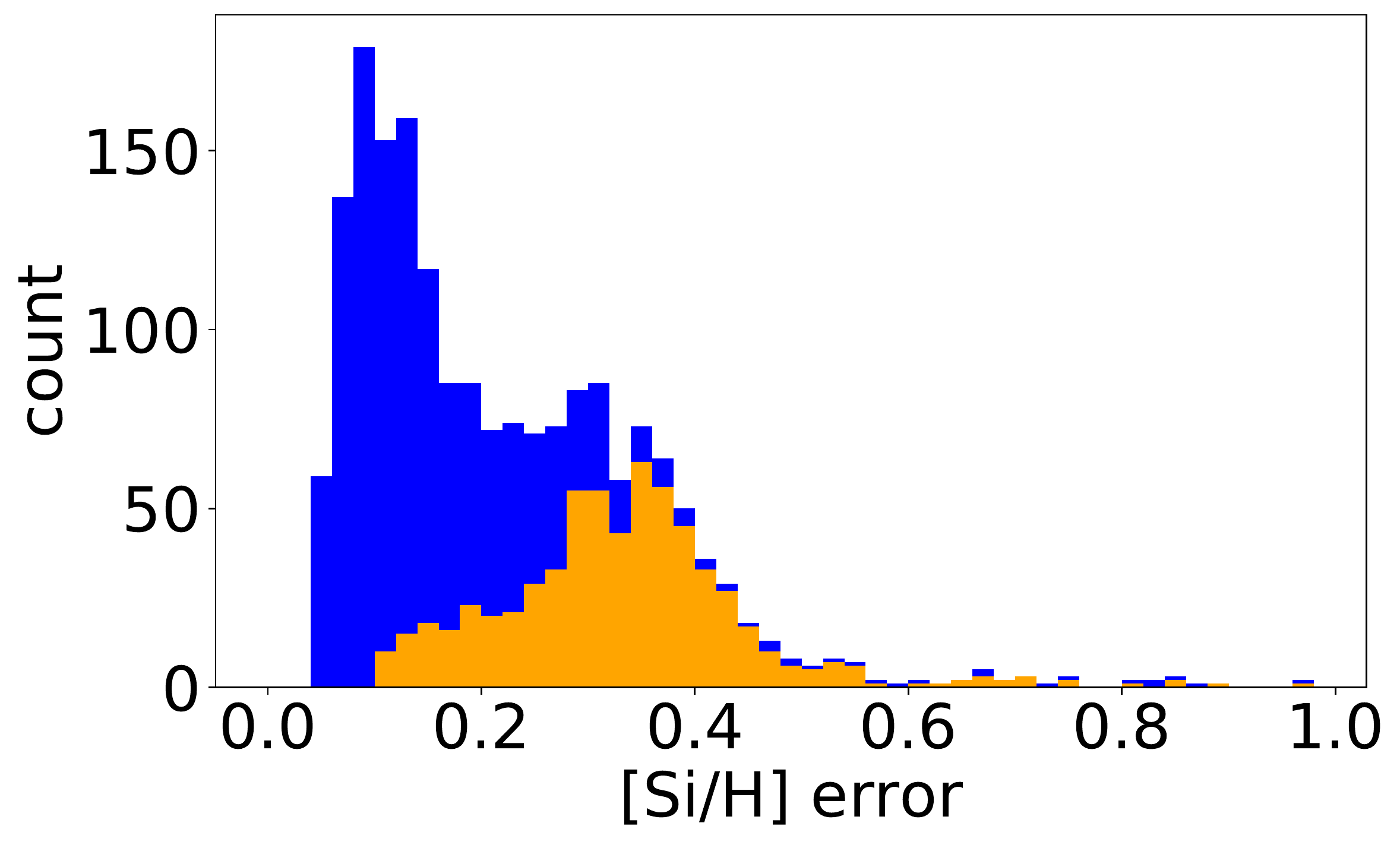}
\includegraphics[width=0.335\linewidth,angle=0]{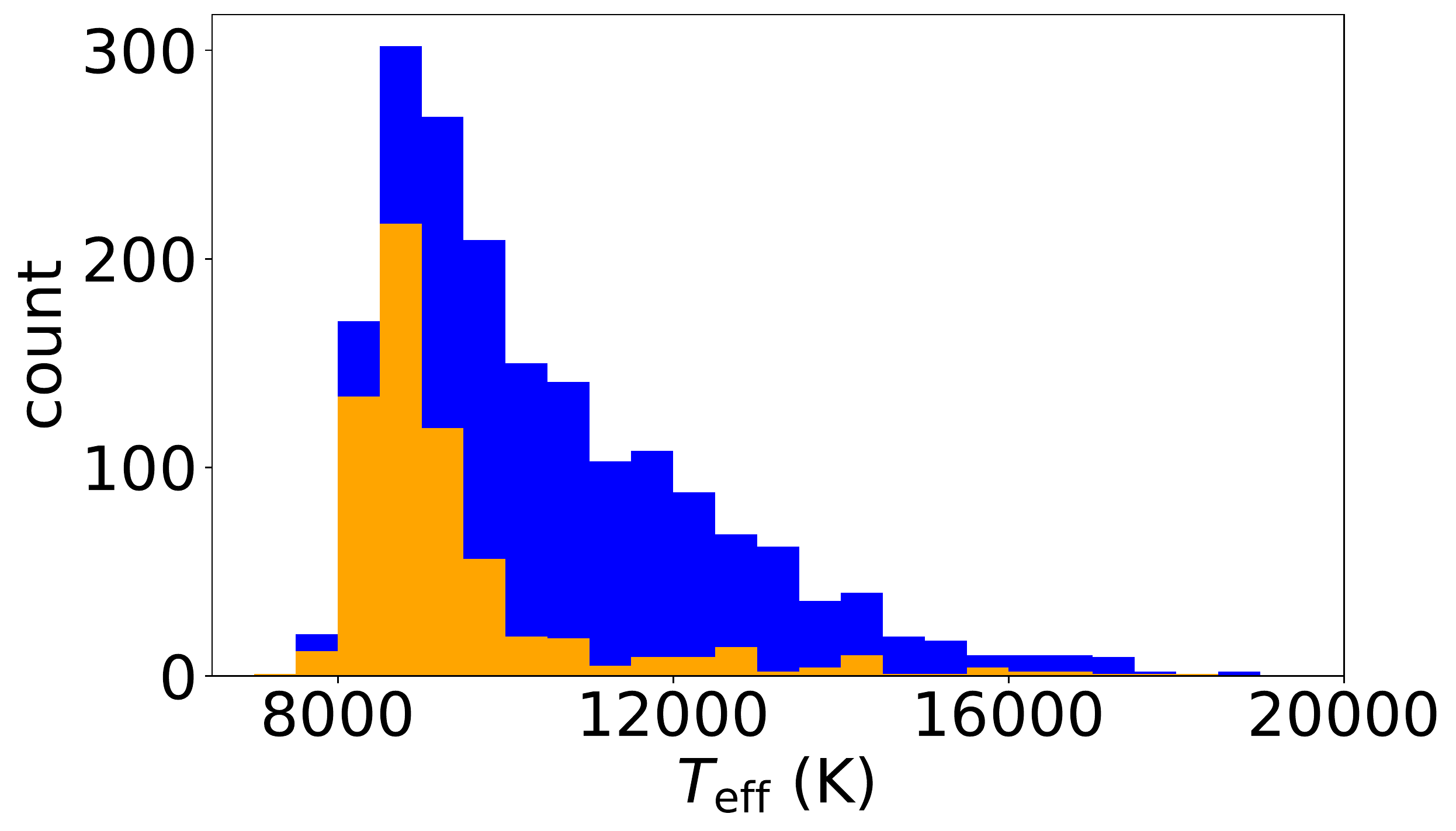}
\includegraphics[width=0.325\linewidth,angle=0]{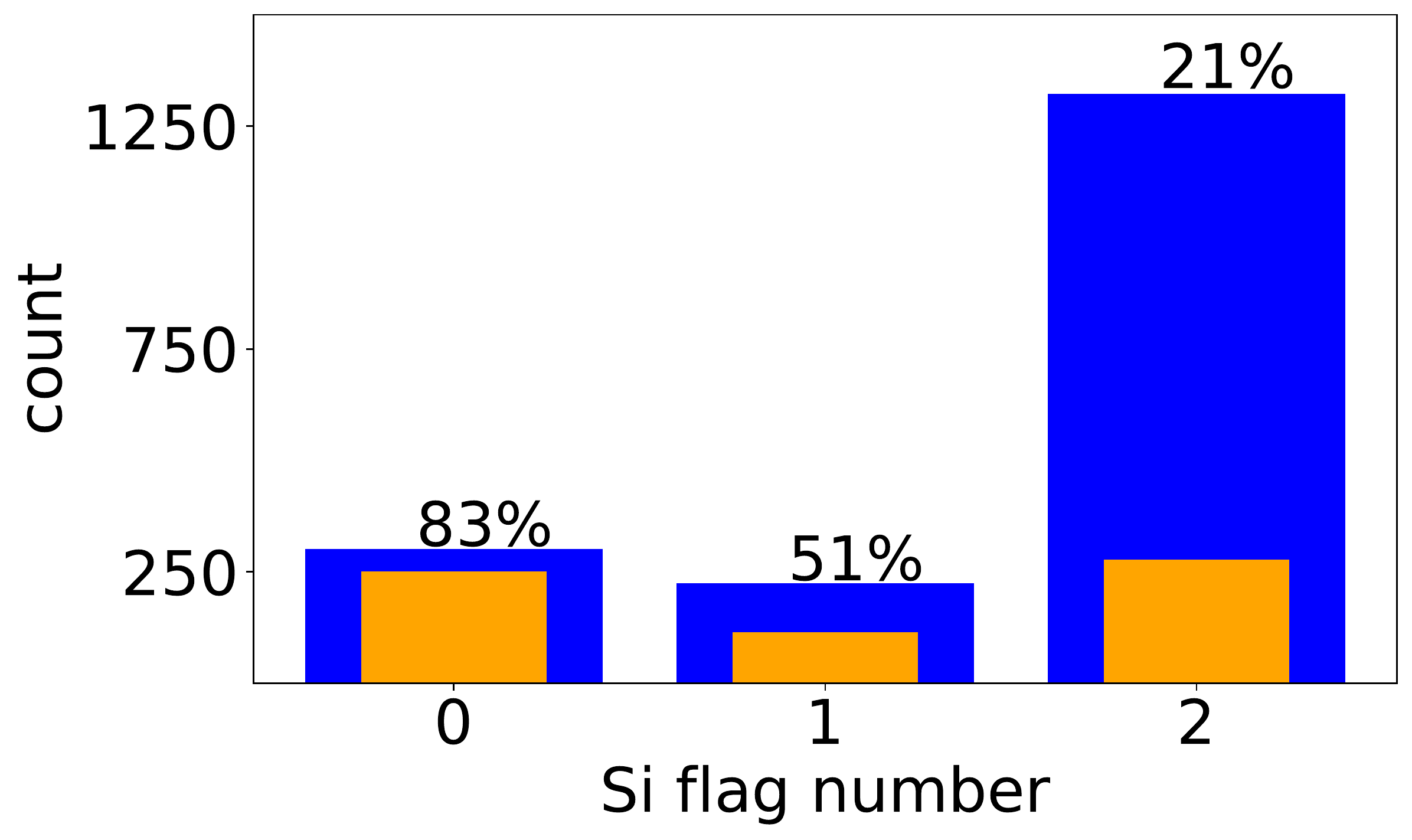}	
\caption{The histogram of [Si/H] errors (left panel), $T_{\rm eff}$ (middle panel), and Si flag number (right panel) of the normal [Si/H] stars in our catalog (orange) and all the Ap stars in our catalog (blue). The fractions of the normal [Si/H] stars among the Ap stars in each group are noted.}
\label{fig:lowsi}
\end{figure}

In the left panel of Fig.~\ref{fig:plane}, [Fe/H] shows normal (abundance lower than 0.2 dex) to deficient Fe for cool Ap stars. This is consistent with the findings of \citet{1996A&A...310..872A} and \citet{mh} and explains why our method tends to select hot stars. In the middle panel, some stars appear to have normal [Fe/H] and [Si/H]. 
First we check the errors of the [Si/H] measurements of these normal [Si/H] stars and compare them with those of other Ap stars in our catalog. 
From the left panel of Fig.~\ref{fig:lowsi}, the distribution of the errors of these normal [Si/H] Ap stars tends to be larger. Moreover, as the Fe abundance in Ap stars goes from underabundant to overabundant as a function of increasing $T_{\rm eff}$, which is shown not only in this work but also in \citet{1996A&A...310..872A} and \citet{mh}, Si abundance may also related with $T_{\rm eff}$. The middle panel of Fig.~\ref{fig:lowsi} shows that the normal [Si/H] stars contain more cooler stars ($T_{\rm eff}<10000$\,K). 
Also, the Ap stars are known to show a large range of abundance peculiarities, with some rare earth elements with nearly normal abundance, so in some cases Si may not be overabundant. 
For example, \citet{2019MNRAS.487.5922G} pointed out that the Si abundances of some Ap stars can even be smaller than the solar abundance.

To verify this, we have checked how many libraries in MKCLASS can give a spectral classification with Si peculiarity for each star (in Section~\ref{pec}). The numbers of libraries which identify Si peculiarity are called Si flag numbers. 
In Section~\ref{pec}, each star is classified by MKCLASS twice comparing with two different libraries. For one star, the Si flag number has three cases: if both of the two libraries identified Si peculiarity, then the Si flag number of this star is 2; if only one library identified Si peculiarity, then the Si flag number of this star is 1; and, if none identified Si peculiarity, then the Si flag number of this star is 0.

The right panel of Fig.~\ref{fig:lowsi} shows the comparison of the Si flag number of normal Si Ap stars and other Ap stars in our catalog. Most of the Ap stars with Si flag number less than 2 are the normal Si Ap stars. Thus, a large error and low $T_{\rm eff}$ may result in an Ap star with normal [Si/H] and [Fe/H] so that there is no conspicuous Si peculiarity in this star. 

Another problem is that some stars with high [Si/H] and [Fe/H] are not selected as Ap stars here. As mentioned before, our method has a bias towards Ap stars with depressions around 5200\,\AA.

\subsection{Evolutionary stage from {\it Gaia}}
\label{gaiaparam}
\begin{figure}
\centering
\includegraphics[width=0.6\linewidth,angle=0]{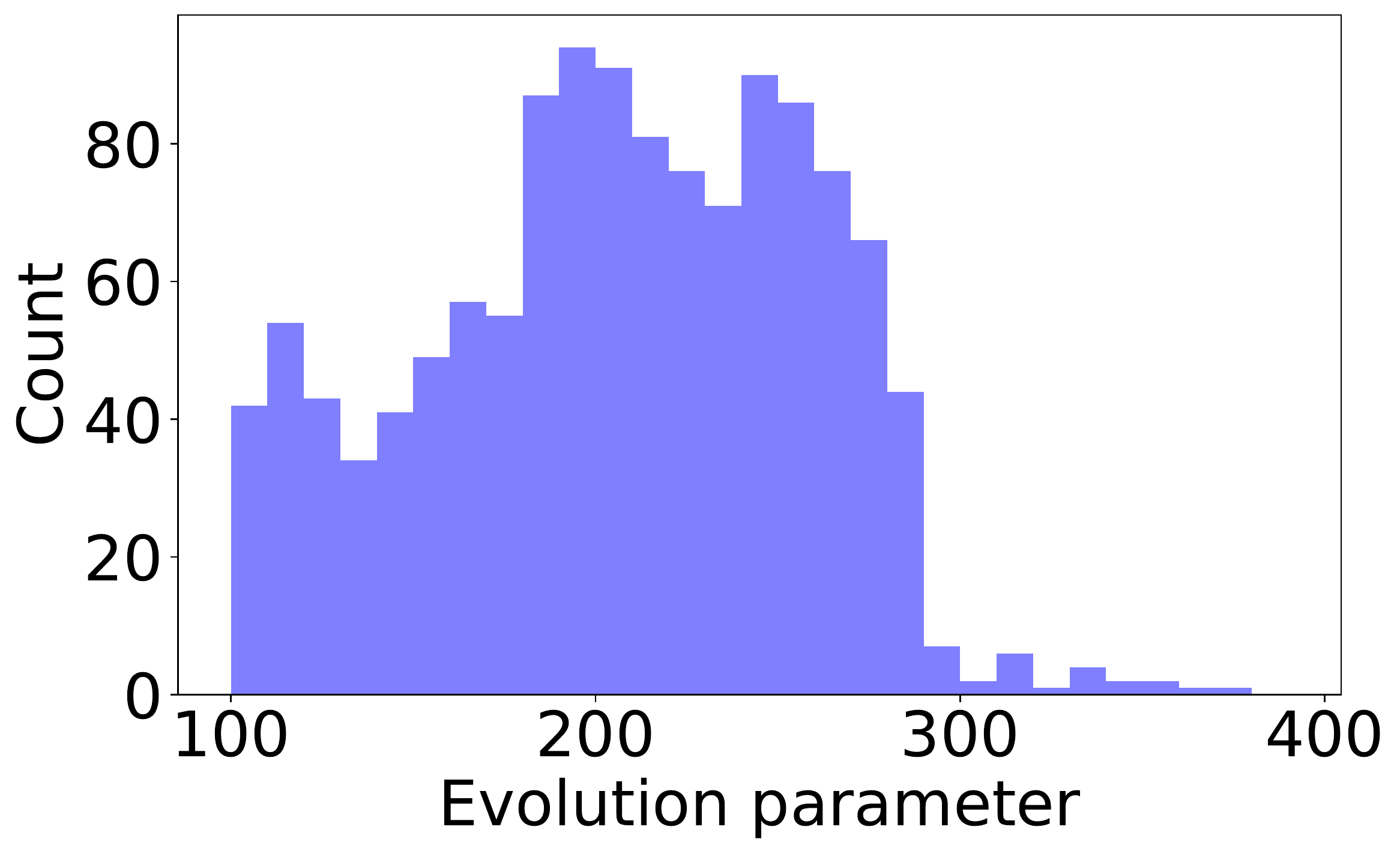}
\caption{Evolutionary stage distribution of 1263 stars with evolution parameters. The x-axis is the evolution parameter, which for main-sequence stars ranges between 100 and 420.}
\label{fig:gaia-evol}
\end{figure}

{\it Gaia} DR3 also provides evolution parameters including mass, age, and evolution stage from FLAME \citep[The Final Luminosity Age Mass Estimator,][]{2022arXiv220605864C} by using  $T_{\rm eff}$, $\log g$, and [M/H] together with distance and magnitude as input parameters to derive luminosity and then comparing with the BASTI \citep{2018ApJ...856..125H} solar metallicity stellar evolution models. Here, we check the evolutionary stage distribution of these stars. In Fig.~\ref{fig:gaia-evol}, all of these stars fall in the range of the main-sequence stage, which is consistent with typical Ap stars and supports our earlier assumption in Section~\ref{teff}

\section{Photometric features}
\label{pho}

Our Ap stars were cross-matched with the target lists of TESS short cadence and {\it Kepler} observations. Forty-two stars (1.6$\%$ of the whole sample) have only TESS short cadence data while 25 stars (0.9$\%$) have only {\it Kepler} data, and five stars (0.2$\%$) have both. To search for rotation and pulsation features, a Fourier transform is applied for each star in both low frequency range (0-60\,d$^{-1}$) and high frequency range (60-300\,d$^{-1}$). All the light curves and the periodograms are inspected by eyes to make double check.

\subsection{Rotation features}

Among these 72 stars, four stars (TIC~2934856, TIC~239801694, TIC~291258469, TIC~434372218) do not show detectable rotation features. \citet{2020A&A...639A..31M} and \citet{2022arXiv220103940M} found a total of 127 super slowly rotating Ap stars (ssrAp stars) which have long rotation periods so that no obvious rotational variability is seen in one TESS Sector, or more. Among our four stars without rotation features detected, TIC~239801694 was identified as a ssrAp star candidate in \citet{2022arXiv220103940M}, and the other three stars may also be candidates or their rotation inclination angles are quite small.

\begin{figure}
\centering
\includegraphics[width=0.3\linewidth,angle=0]{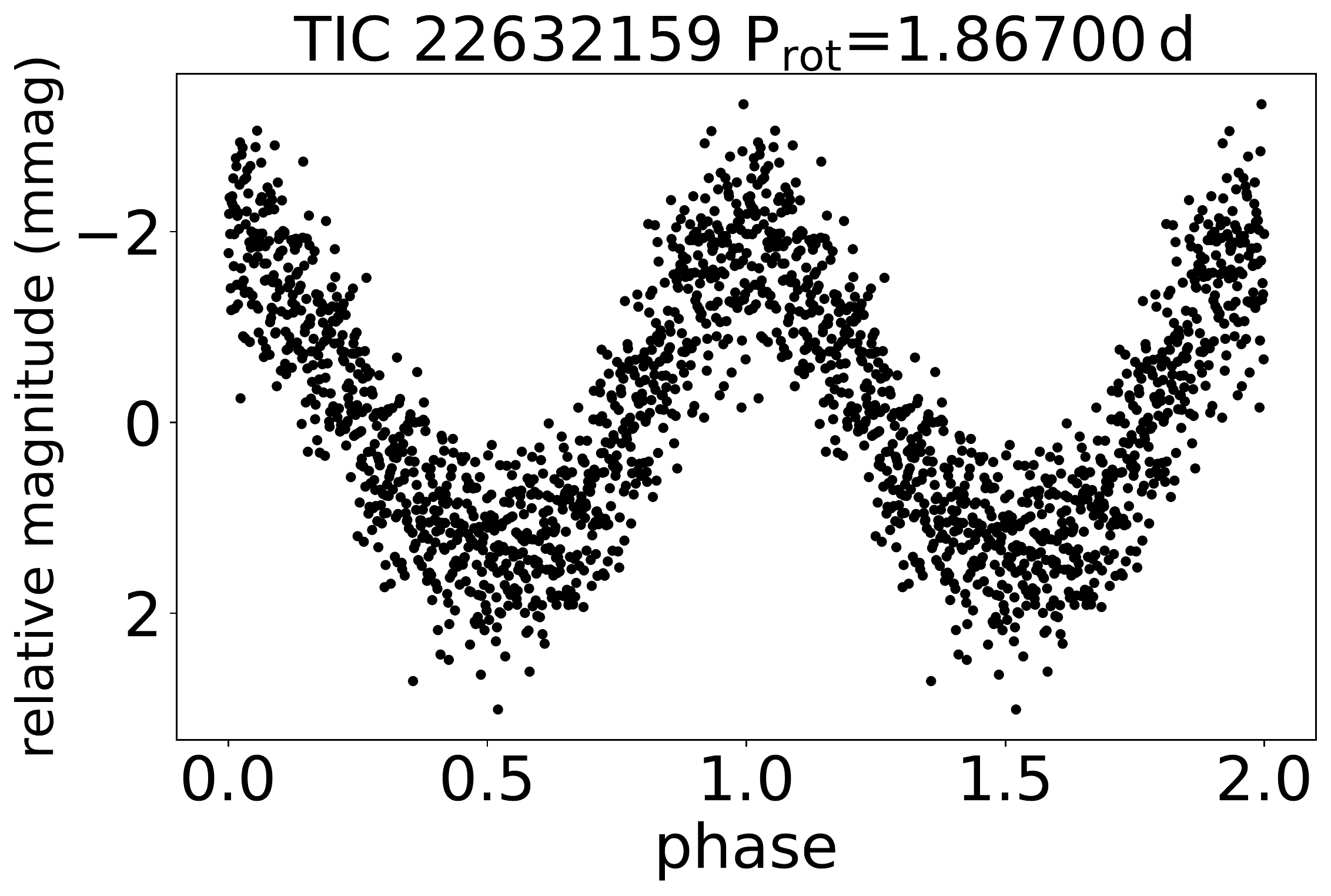}
\includegraphics[width=0.3\linewidth,angle=0]{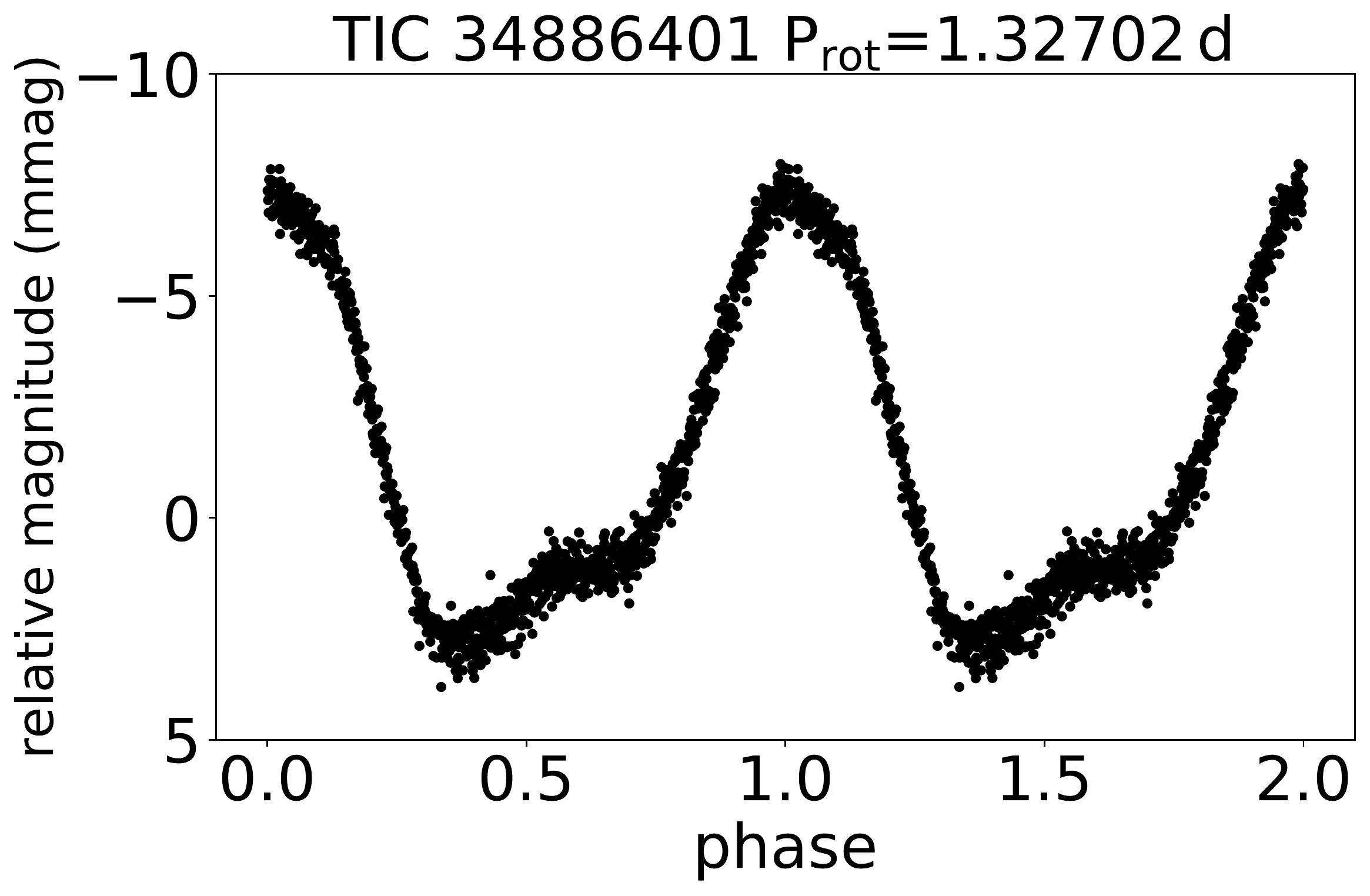}
\includegraphics[width=0.3\linewidth,angle=0]{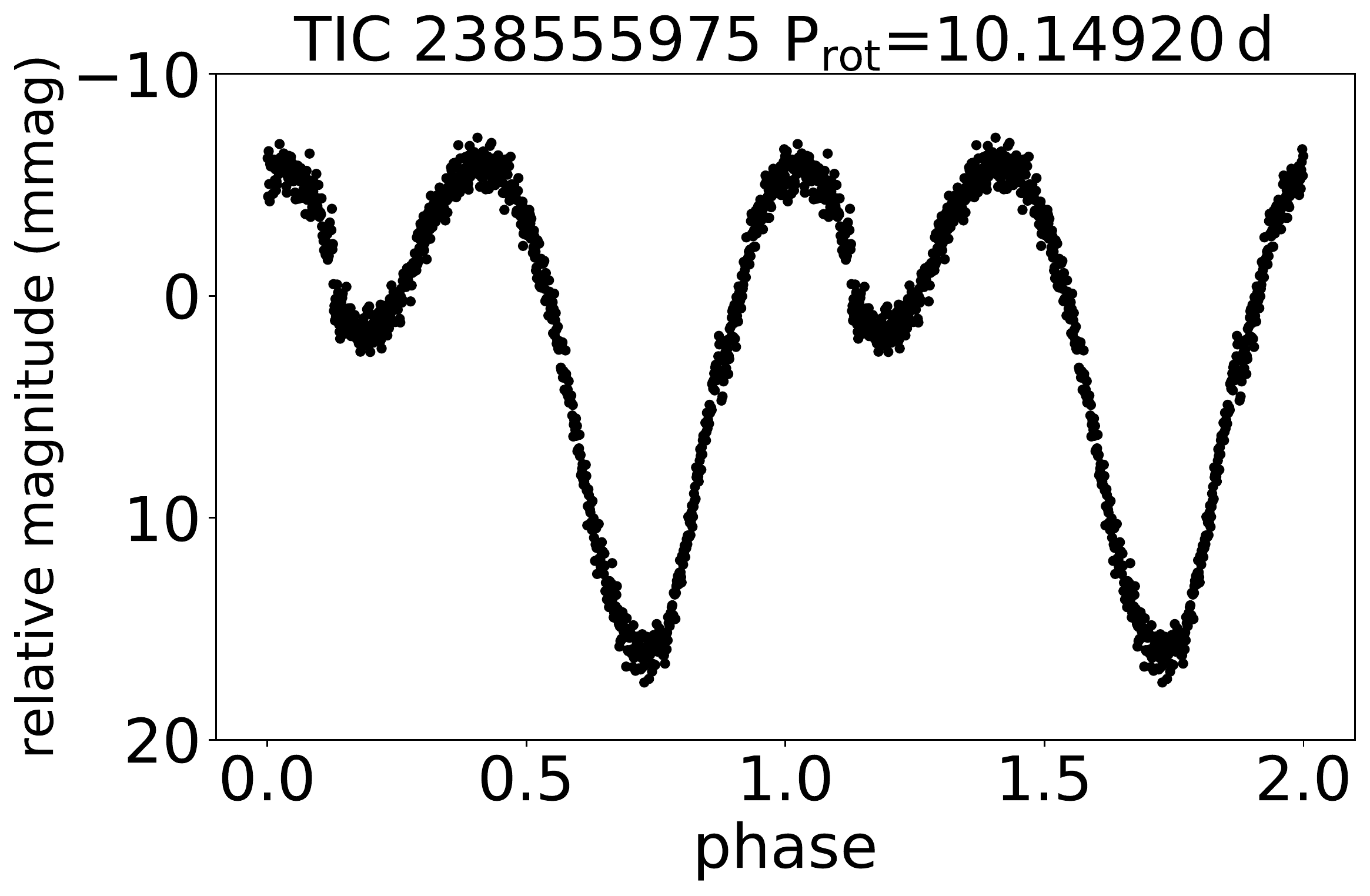}
\includegraphics[width=0.3\linewidth,angle=0]{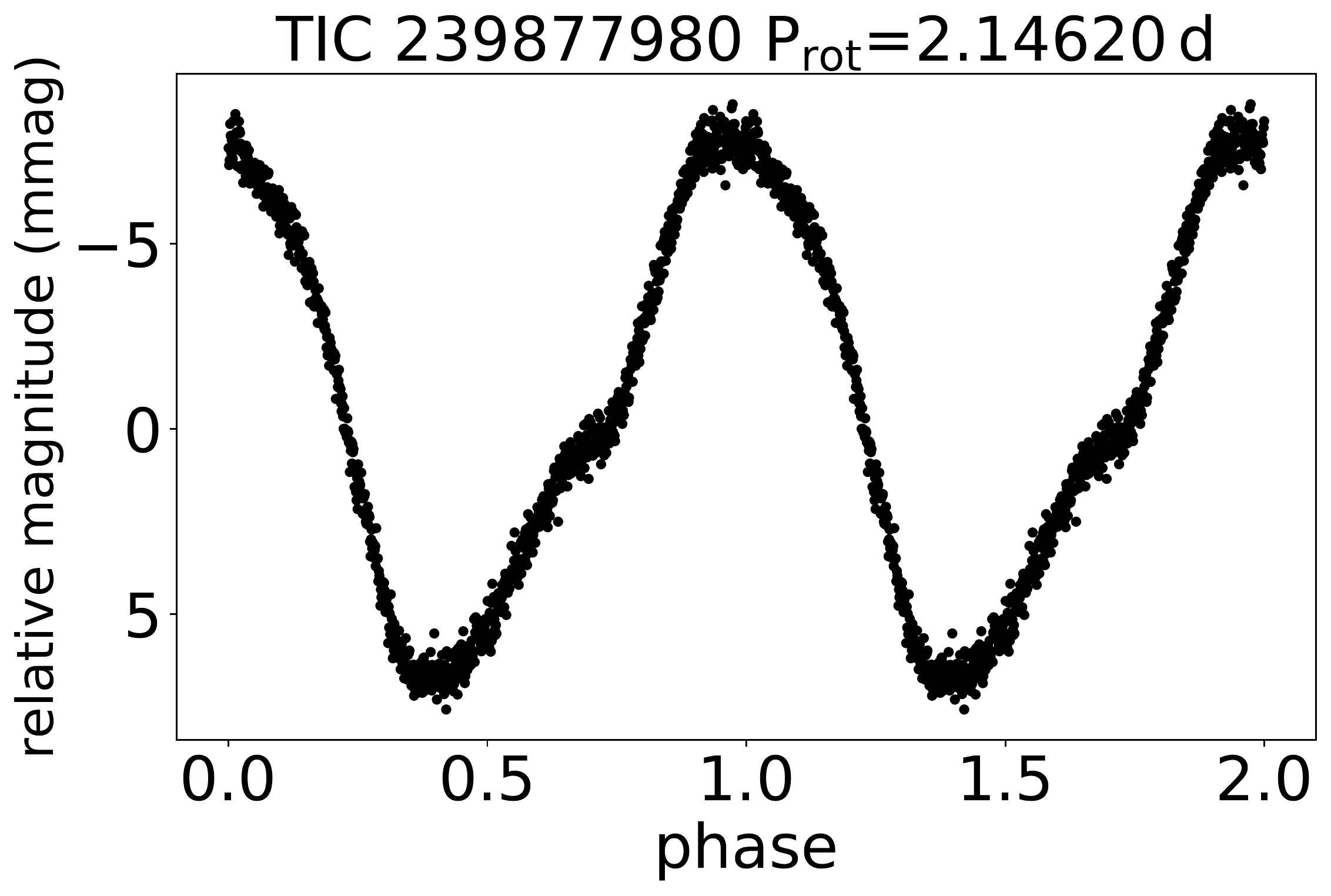}
\includegraphics[width=0.31\linewidth,angle=0]{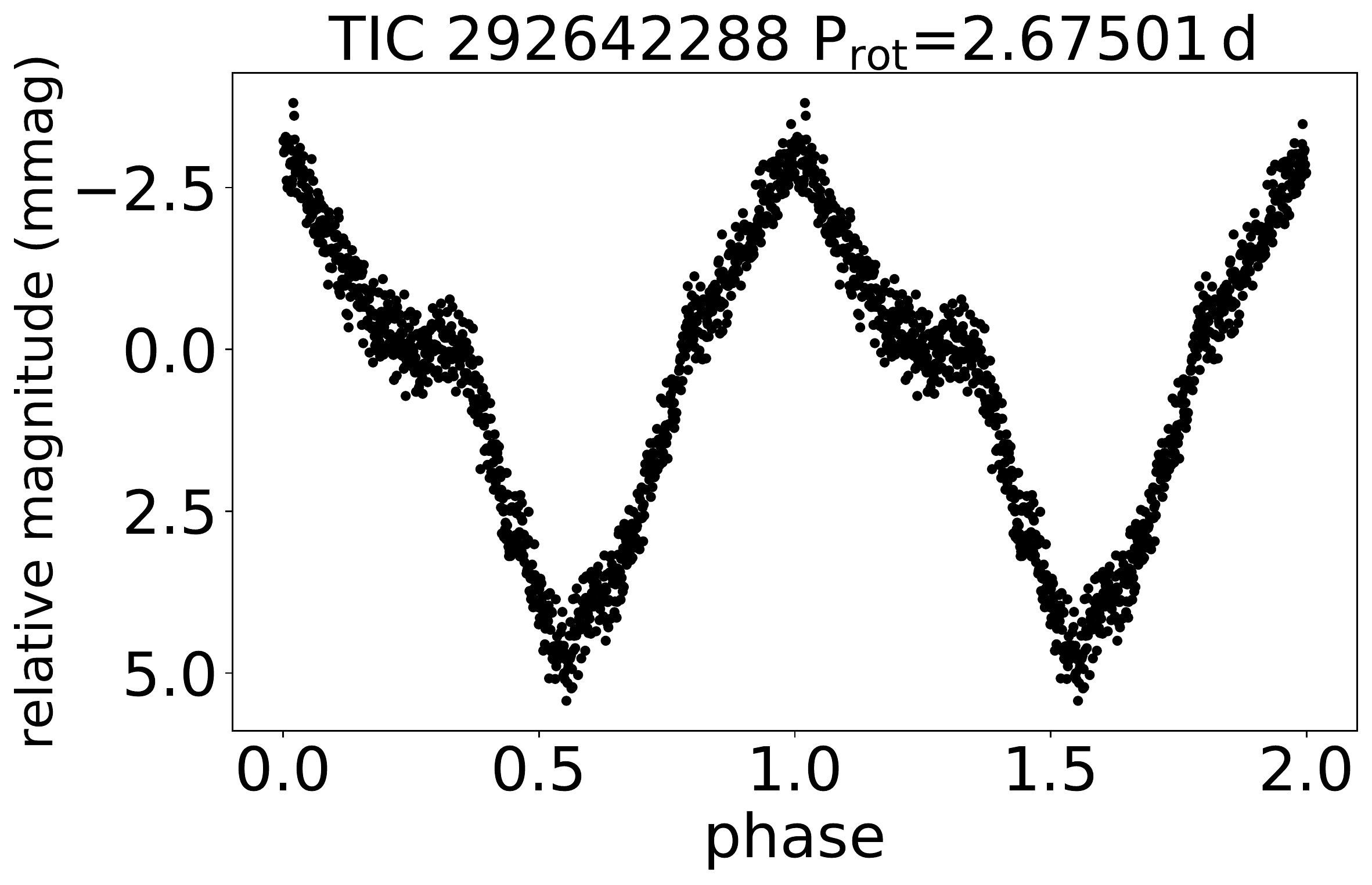}
\includegraphics[width=0.31\linewidth,angle=0]{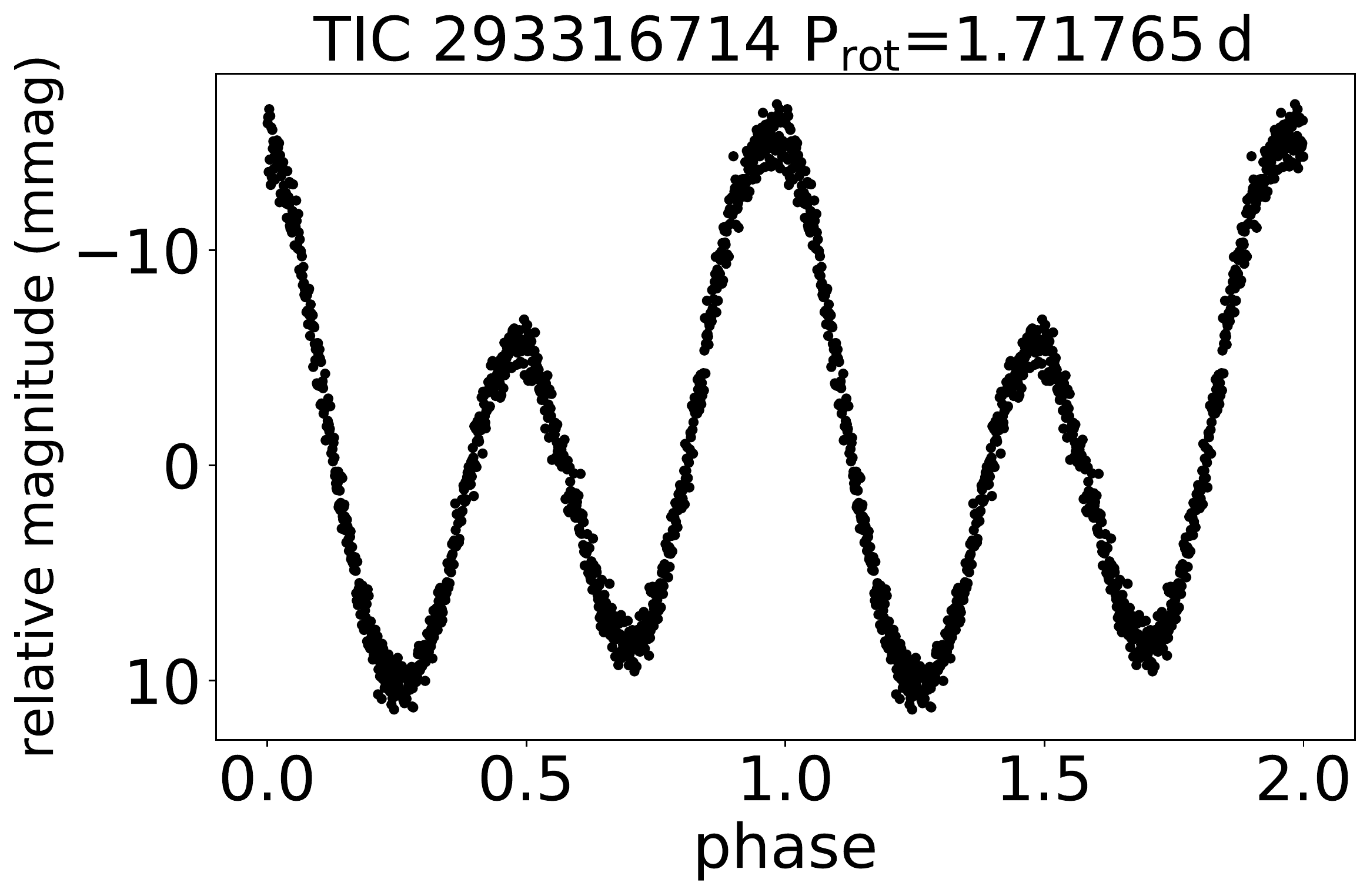}
\includegraphics[width=0.3\linewidth,angle=0]{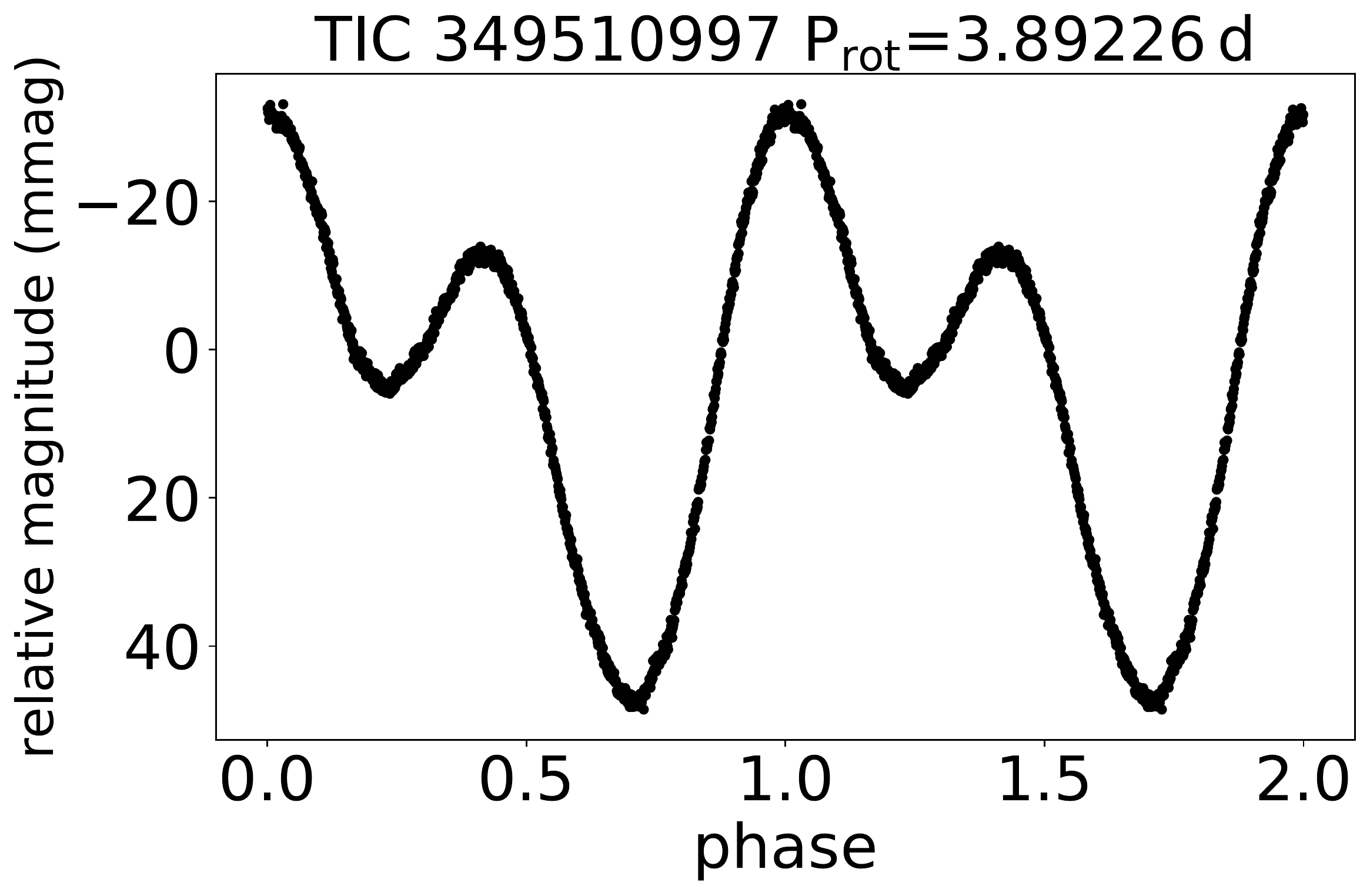}	

\caption{The phase folded light curves of new rotational variable stars. The TIC IDs and periods are shown in each title.}
\label{fig:lc_rot}
\end{figure}

Seven stars (TIC~22632159, TIC~34886401, TIC~238555975, TIC~292642288, TIC~293316714, TIC~239877980, TIC~349510997) show clear rotation features but they are not included in any related literature. These are new rotational variable stars. Their periods and phase folded light curves are given in Fig.~\ref{fig:lc_rot}. In addition, seven other stars (TIC~48354181, TIC~83096510, TIC~121603913, TIC~121732964, TIC~273130000, TIC~274024023, TIC~416528957) were identified as eclipsing binaries in the literature and they are noted in the catalog. The spectra of these seven binary stars are visually checked to make sure that they have Ap star peculiarities like enhancement of Si, Eu, and Sr. Although the binary systems mostly appear in Am stars whereas only about 20\% Ap stars are in binary systems \citep{1973ApJS...25..137A}, the possibility that these stars are Ap stars cannot be ruled out either.

\subsection{Pulsation features}
For pulsation features, we find one roAp star candidate, TIC~21024812, one $\delta$ Scuti star candidate, TIC~119253179, and a known roAp star, TIC~272598185, whose roAp pulsation decreased and a new $\delta$ Scuti pulsation appeared. These stars will be discussed in the following subsections. 

\subsubsection{The new roAp star, TIC~21024812 (HD~14522)}

\begin{figure}
\centering
\includegraphics[width=0.4\linewidth,angle=0]{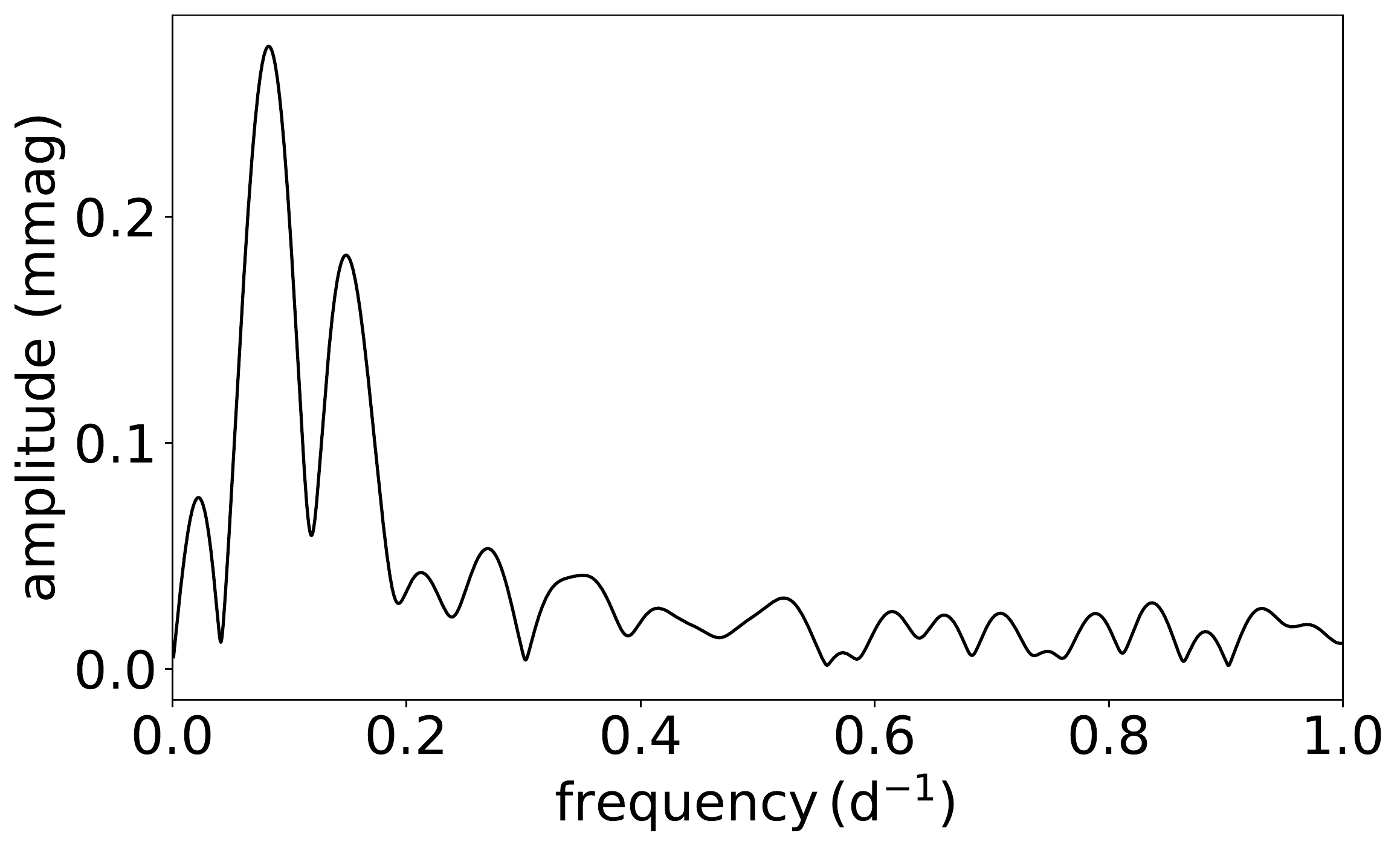}
\includegraphics[width=0.42\linewidth,angle=0]{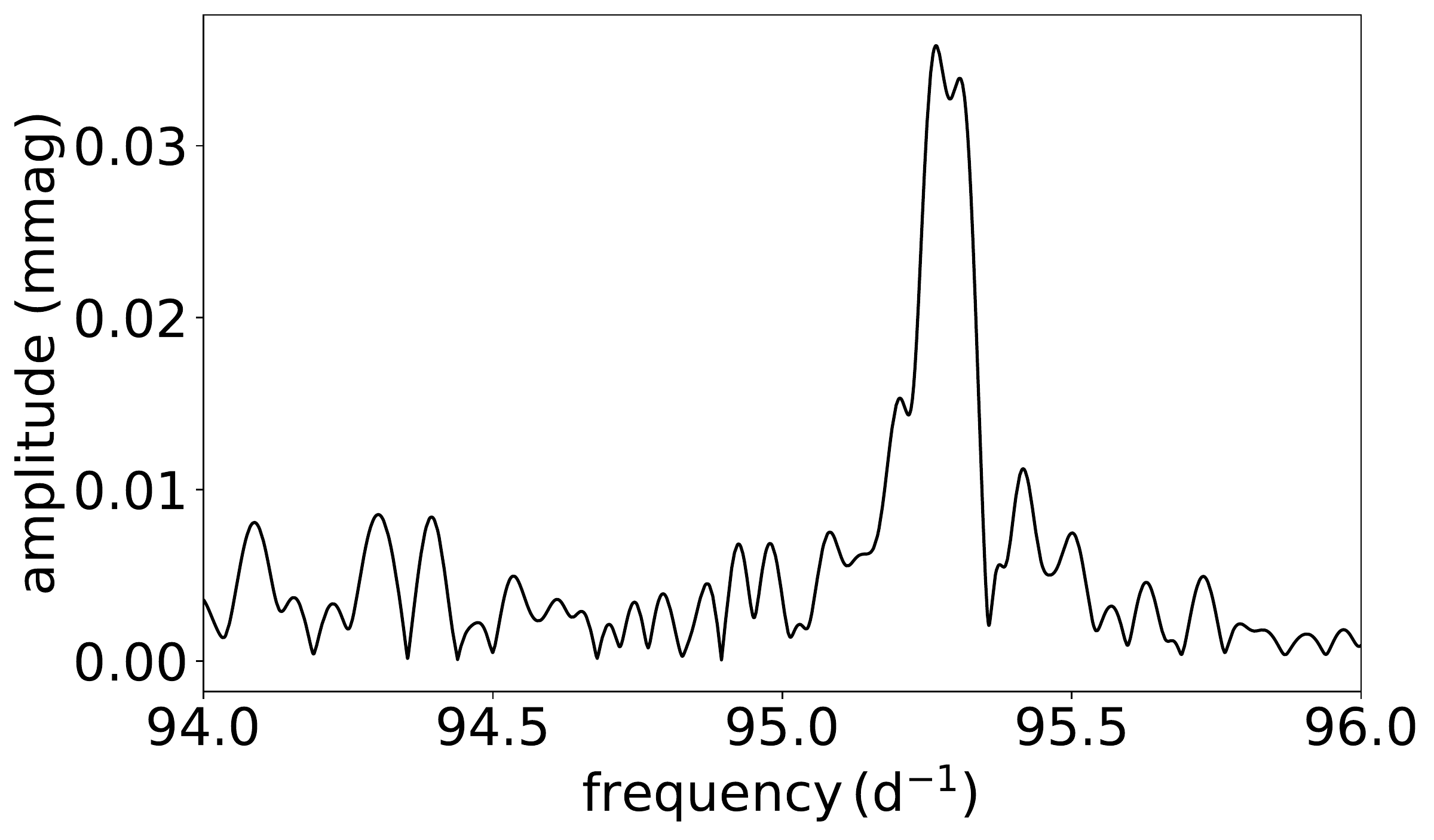}
\caption{The periodogram of TIC~21024812 in low- and high-frequency range.}
\label{fig:ft_hd14522}
\end{figure}

TIC~21024812 was identified as an Ap star by \citet{1991A&AS...89..429R} with the spectral type of A2 SrEu. It has been discussed in several works (e.g. \citealt{2006A&A...455..303J,2020MNRAS.493.3293B}) which also searched for pulsations but no detection was reported. This star was observed in TESS Sector~18. A Fourier transform shows a significant low-frequency peak at 0.07174\,d$^{-1}$ (left panel in Figure~\ref{fig:ft_hd14522}). Taking the rotation frequency to be half of this, i.e., there is a double wave rotational variation, gives a rotation frequency of $\nu_{\rm rot} = 0.03587$\,d$^{-1}$ ($P_{\rm rot} = 27.88$\,d) which is comparable to the rotation period given in \citet{2020MNRAS.493.3293B}, $P_{\rm rot} = 26.36$\,d. As the observation time span, 24\,d, is shorter than the rotation period, and there are gaps in this sector of data, the pipeline processing may suppress some low frequencies hence one cannot get an accurate rotation period.

After a Fourier transform in the high frequency range, we have found a principal pulsation frequency at $95.264$\,d$^{-1}$, which is not fully resolved, as can be seen in Fig.~\ref{fig:ft_hd14522} (right panel). To get optimised result of frequencies, linear and non-linear least squares fits are used after removing rotation and other low frequency signals. Since the gaps in TESS data make it difficult to determine the rotation period, we use the rotation period (P=26.36\,d) and the epoch (BJD=2454059.3) given by \citet{2020MNRAS.493.3293B} to set the zero point time to be t$_0$ = BJD~2458803.96, which is also near the middle time of the observed data. The result of frequency analysis is listed in Table~\ref{Tab:freq_hd14522}. A pre-whitening process leads to one side-lobe separated from the main frequency due to the rotation frequency, hence suggests oblique pulsation. However, given the resolution limitation, this result is not robust. A longer time span of data is needed to confirm the rotation frequency.

\begin{table}
\scriptsize
\centering
\caption{A non-linear least squares fit of the frequency multiplets for TIC~21024812. The zero point for the phases is t$_0$ = BJD~2458803.96.} 
\setlength{\tabcolsep}{1cm}{
\begin{tabular}{llll}
\hline
\hline
 & frequency & amplitude & phase \\
 & (d$^{-1})$ & (mmag) & (radians) \\
\hline
$\nu_1$ & 95.263 $\pm$ 0.002 & 0.038 $\pm$ 0.003 & 2.37 $\pm$ 0.10 \\
$\nu_2$ & 95.311 $\pm$ 0.002 & 0.036 $\pm$ 0.003 & 1.65 $\pm$ 0.10 \\

 \hline
 \hline
\end{tabular}}
\label{Tab:freq_hd14522}
\end{table}

\subsubsection{The new $\delta$ Scuti pulsation of TIC~272598185 (KIC~10483436)}

TIC~272598185 was identified as a roAp star with the frequency around 117\,d$^{-1}$ by \citet{2011MNRAS.413.2651B} using {\it Kepler} short cadence data. 
Additionally, this star was observed 10 years after {\it Kepler}'s observations in TESS sectors 14, 15, 41, and 54.

\begin{figure}
\centering
\includegraphics[width=0.45\linewidth,angle=0]{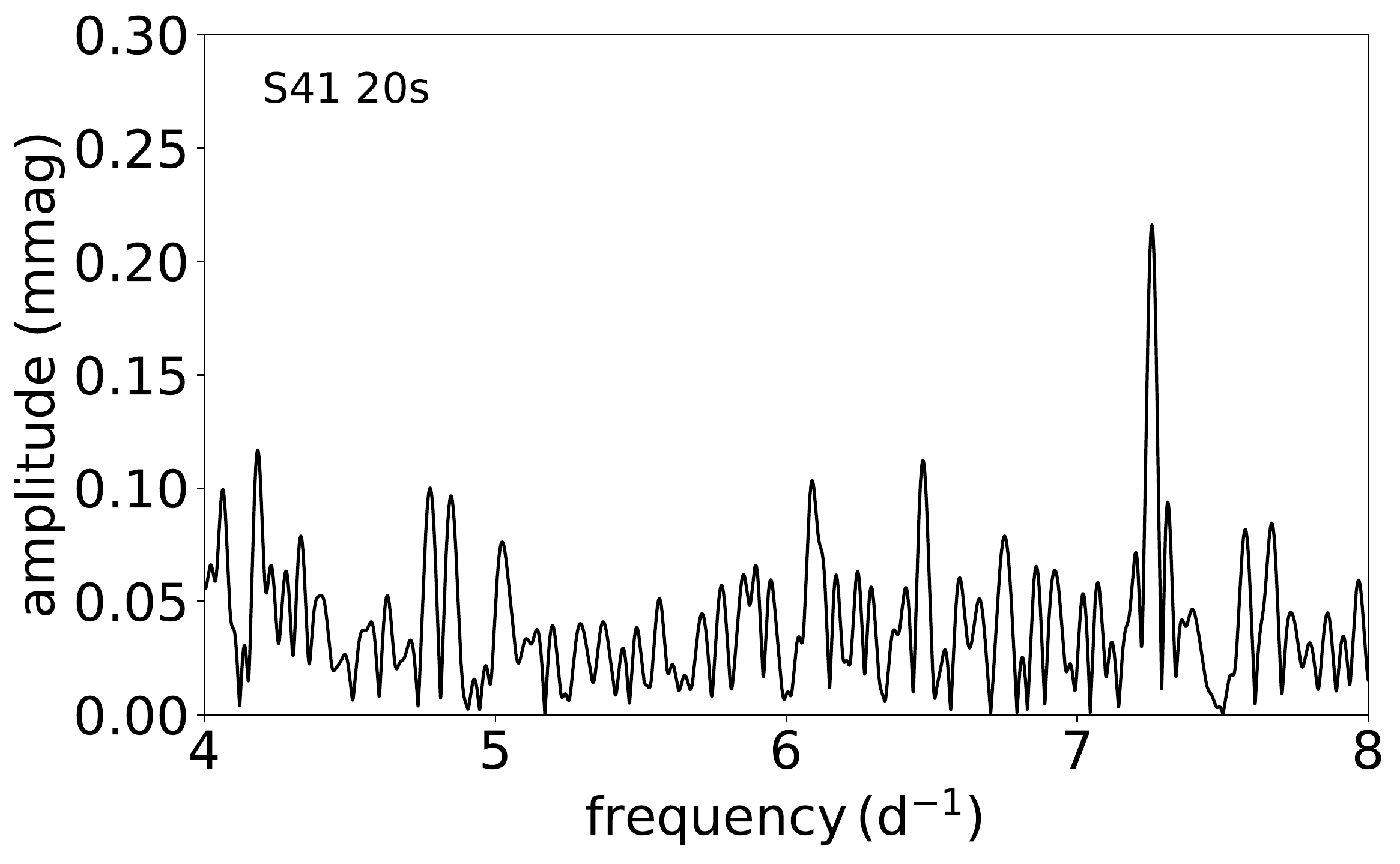}
\includegraphics[width=0.45\linewidth,angle=0]{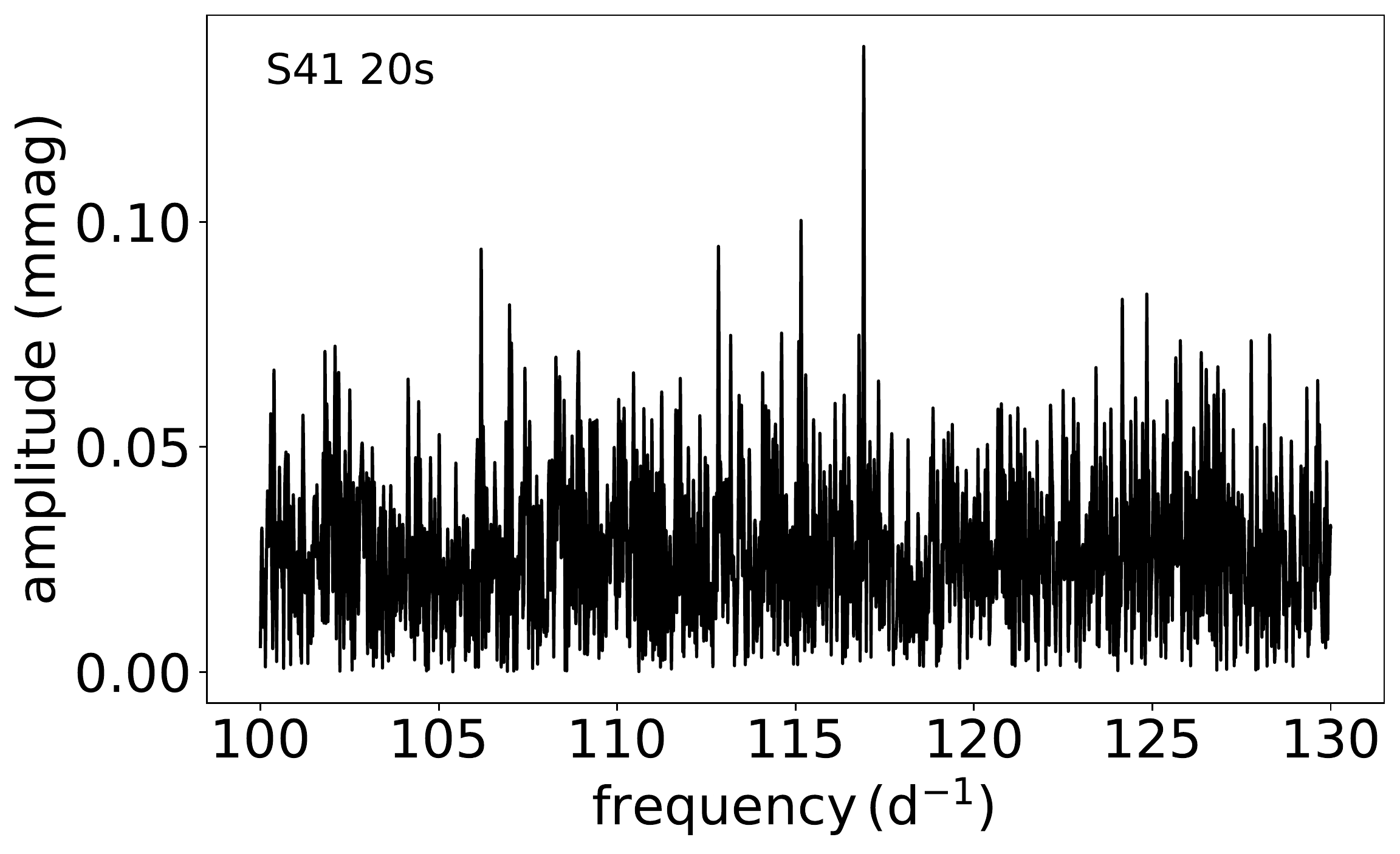}	
\includegraphics[width=0.45\linewidth,angle=0]{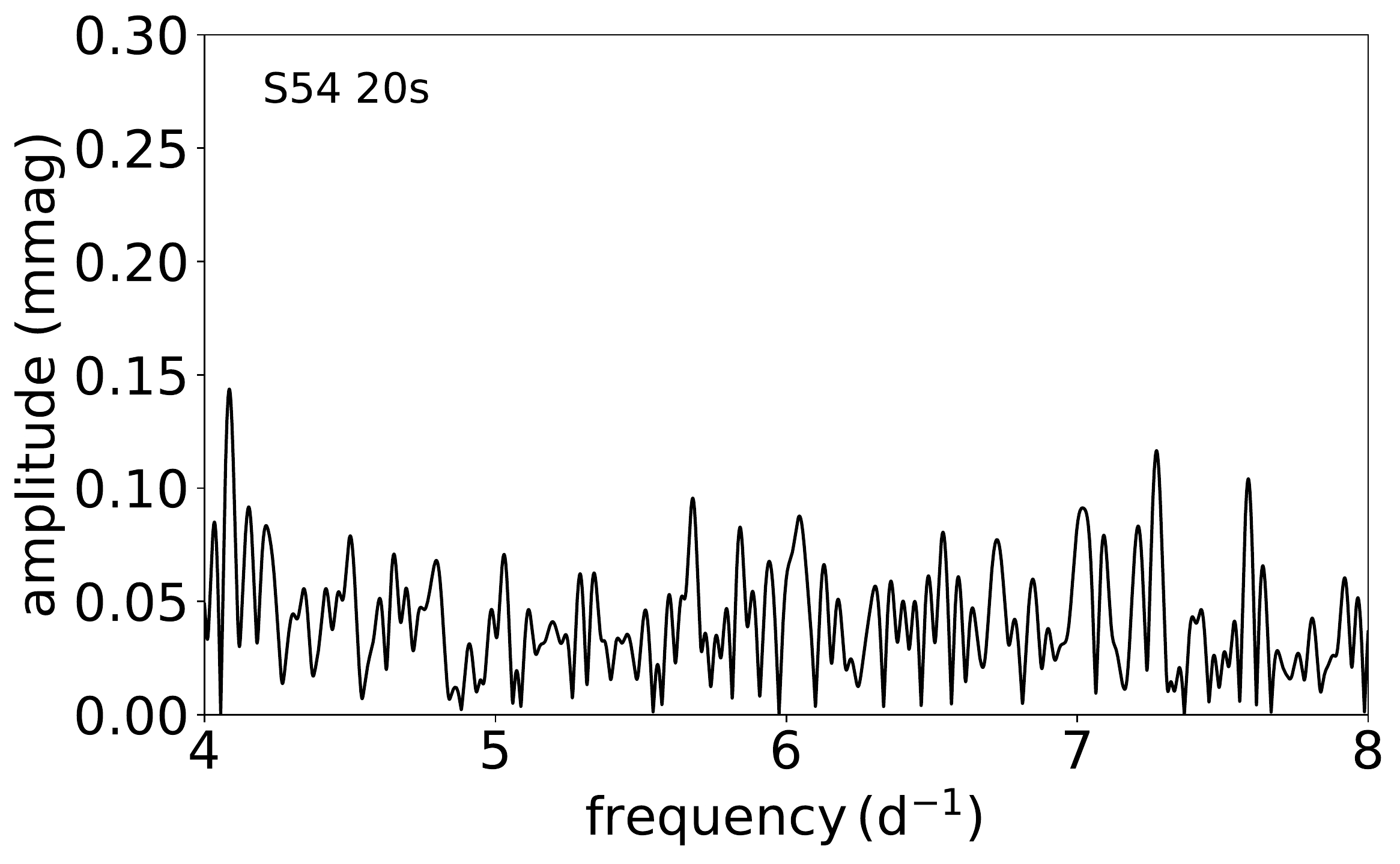}
\includegraphics[width=0.45\linewidth,angle=0]{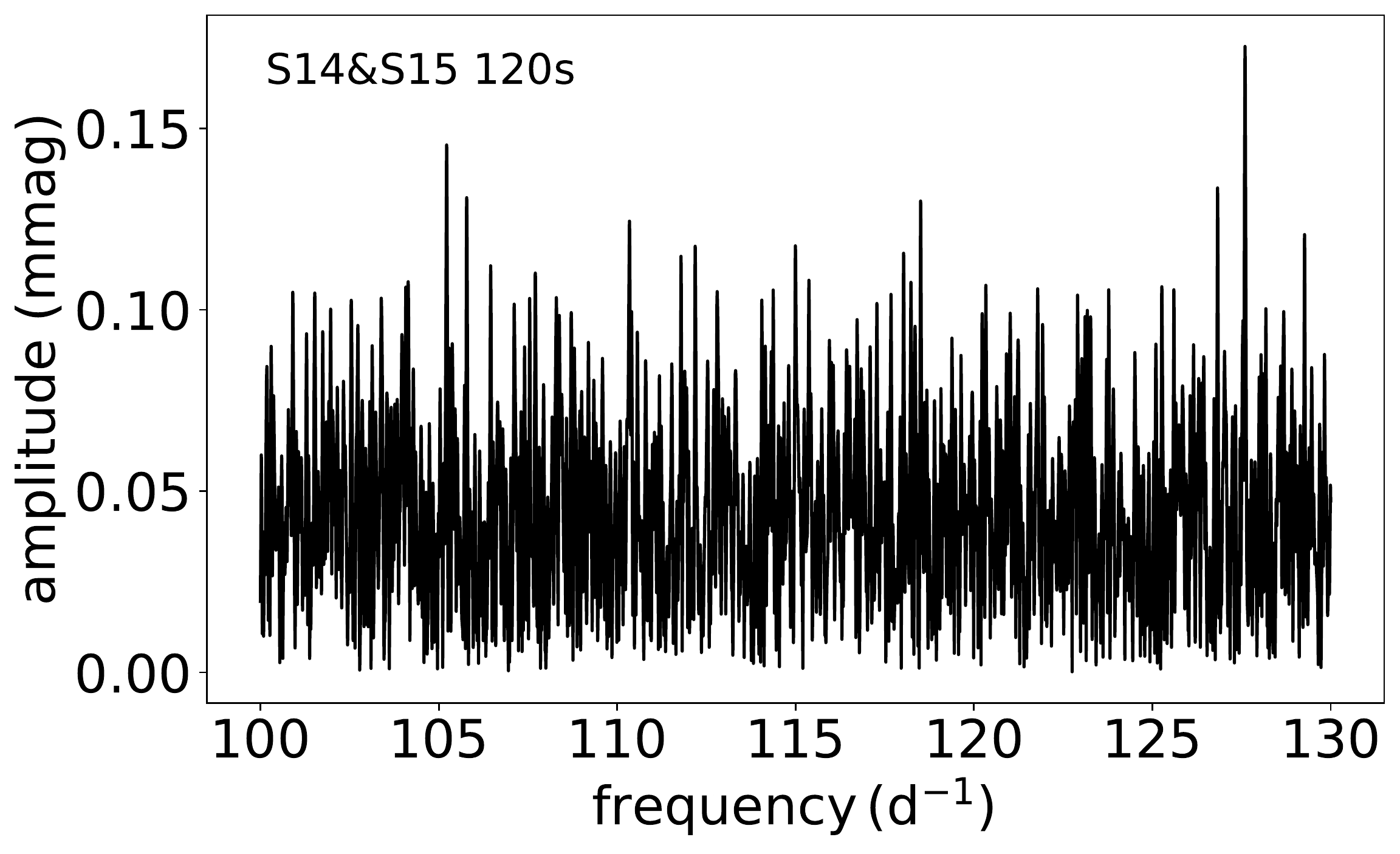}
\caption{The periodogram of TIC~272598185 with the data from TESS sector~41 20\,s cadence in low- and high-frequency range (top two panels), TESS sector~54 20\,s cadence (left bottom), and TESS sector~14 and 15 120\,s cadence (right bottom) zoomed in around 7.25 d$^{-1}$ and 117\,d$^{-1}$ to show significant pulsation signals (top) or null results (bottom).}
\label{fig:ft_kic10483436}
\end{figure}

The roAp pulsation (right top panel of Figure~\ref{fig:ft_kic10483436}) around 117\,d$^{-1}$ is marginally detected in the Sector 41 20-s data with an amplitude just above 100\,mmag which is close to the noise level. But it falls lower than the high noise in the sector~14 and 15 120-s data. The amplitude detected by TESS is above 0.1\,mmag which is significantly larger than the 0.07\,mmag from {\it Kepler}. This may be caused by the slightly different passbands of the two missions, or by a change in amplitude in the 10 years between the data sets. 

A newly detected significant $\delta$ Scuti type pulsation around 7.25 d$^{-1}$ can be seen in the Sector 41 20\,s data (Figure~\ref{fig:ft_kic10483436}, left top panel), but not the Sector~54 data (Figure~\ref{fig:ft_kic10483436}, left bottom panel). The presence of the low-amplitude $\delta$~Sct mode suggests a probable magnetic field strength less than a few kG \citep{2020MNRAS.498.4272M}.

\section{Conclusions}
\label{conclude}

We have found 2700 Ap stars by searching LAMOST DR9 for the 5200\,\AA\ flux depression, with all of the spectra having been inspected by eyes. Detailed spectral classification of our Ap stars are given by applying a modified MKCLASS program which was developed by HPB2020.

Our Ap star catalog is compared with other related catalogs. Since each catalog searched for CP stars by different methods, the stars identified as Ap star candidates may have different selection bias leading to few stars in common among these catalogs. 

Statistical analyses of the stellar parameters $T_{\rm eff}$, $\log g$, absolute magnitude, mass and evolutionary status are carried out. We find that most stars have $T_{\rm eff}$ and $\log g$ typical of main sequence stars with masses in the range of $2-3$\,M$_\odot$. These properties are in agreement with those of previously known Ap stars. From the evolution parameters given by {\it Gaia} DR3, all the stars are in main sequence.

The metallicities, [Si/H] and [Fe/H], together with $v \sin i$ of our Ap stars are compared with those of non-peculiar stars. Lower $v \sin i$, higher Fe abundance, and much higher Si abundance of our Ap stars can be clearly seen. 

After crossmatching our Ap star catalog with the target lists of TESS and {\it Kepler} data, we inspect the rotation and pulsation features of 72 stars in common. Among these Ap stars, seven stars are new rotational variable stars and three ones do not show rotation features hence they may be super slowly rotating Ap star candidates. We also find a new roAp star, TIC~21024812. For a known roAp star, TIC~272598185, we find a new $\delta$ Scuti pulsation appeared.

Our Ap star catalog is complementary to other existing Ap star catalogs, and the method to search for Ap star candidates was shown to be valid. Our catalog is hopefully valuable for the further study of Ap stars. In the future, by measuring the magnetic fields and obtaining high-resolution spectroscopic observations of the Ap stars in our catalog, the effect of the magnetic field on the abundance and rotation can be studied. Combining with the photometric data, these stars are also good candidates for searching for roAp stars and ssrAp stars.

\section*{Acknowledgements}
This work was funded by the National Natural Science Foundation
of China (NSFC Grant No.11973001, 12090040, 12090044, U1731108 and 11833002) and the National Key R$\&$D Program of China (No. 2019YFA0405500).

This work has made use of data products from the Guo Shou Jing Telescope (the Large Sky Area Multi-Object Fibre Spectroscopic
Telescope, LAMOST). LAMOST is a National Major Scientific
Project built by the Chinese Academy of Sciences. Funding
for the project has been provided by the National Development and
ReformCommission. LAMOST is operated and managed by theNational
Astronomical Observatories, Chinese Academy of Sciences.

We used data from the European Space Agency mission
{\it Gaia} (http://www.cosmos.esa.int/{\it Gaia}), processed by the
{\it Gaia} Data Processing and Analysis Consortium (DPAC; see
http://www.cosmos.esa.int/web/{\it Gaia}/dpac/consortium).
\\
\\
Some of the data presented in this paper were obtained from the Mikulski Archive for Space Telescopes (MAST) at the Space Telescope Science Institute. The specific observations analyzed can be accessed via \dataset[10.17909/j98f-cs20]{https://doi.org/10.17909/j98f-cs20}. 

\appendix
\setcounter{table}{0}
\renewcommand{\thetable}{A\arabic{table}} 
\begin{longrotatetable}
\begin{deluxetable*}{llllcllllll}
\tablecaption{Example of our Ap star catalog. The full catalog is available electronically.\label{Tab:cat}}
\tabletypesize{\tiny}
\tablehead{
\colhead{obsid} & 
\colhead{RA} & \colhead{DEC} & 
\colhead{$G_0$} & 
\colhead{$(BP-RP)_0$} & \colhead{$a$} & \colhead{distance} &
\colhead{spt-liblamost} & \colhead{spt-libnor36} & \colhead{note} & \colhead{{\it Gaia} id} \\ 
 \colhead{} & \colhead{(degree)} & \colhead{(degree)} & 
 \colhead{(mag)} & \colhead{(mag)} &
\colhead{(mag)} & \colhead{{pc}} &
\colhead{} & \colhead{} & \colhead{} & \colhead{} 
} 
\startdata
2416019 & 29.31 & 59.79 & 0.52 & 0.12 & 0.58 & 1721.97 &  kB9.5hA2mA4  bl4077bl4130Si3856Si5041Si5056 &  A0 II-III  bl4077bl4130Si3856Si5041Si5056 &  & 507868307825704320\\
4116106 & 46.79 & 53.86 & -0.09 & 0.02 & 0.58 & 679.40 &  B8 III  (bl4077)bl4130Si3856Si5056Cr4172 &  B8 III-IV  bl4130Si3856Si5041Si5056Cr4172 &  & 447095826254298112\\
8312154 & 52.41 & 6.09 & 1.13 & -0.04 & 0.59 & 873.41 &  B9 IV-V  bl4077bl4130Si3856Si5041Si5056Cr4172 &  B9 IV-V  bl4077bl4130Si3856Si5041Si5056Cr4172 &  & 3276185836320362624\\
15005220 & 55.73 & 48.90 & 0.67 & 0.08 & 0.59 & 2583.10 &  B9 II  (bl4077)bl4130Si5056Cr4172 &  A0 II  (bl4077)bl4130Si5041Si5056 &  & 248775986236536576\\
15010139 & 53.09 & 50.21 & -0.15 & 0.08 & 0.59 & 3099.77 &  B8 III  bl4077bl4130Si5041Si5056Cr4172 &  A0 II-III  bl4077bl4130Si5041Si5056Cr4172 &  & 441988560381708160\\
15016079 & 54.63 & 51.57 & -0.85 & 0.11 & 0.58 & 1920.18 &  B8 III-IV  (bl4077)bl4130Si3856Cr4172 &  B9.5 III  (bl4077)bl4130Si3856Si5056Cr4172 &  & 442169631902357248\\
15016105 & 55.70 & 51.98 & 1.26 & 0.08 & 0.60 & 1275.35 &  B8 IV  bl4077bl4130Si3856Si5041Si5056Cr4172 &  B9 III-IV  bl4077bl4130Si3856Si5041Si5056Cr4172 &  & 443661948360036352\\
15301125 & 34.22 & 56.52 & 0.28 & -0.00 & 0.57 & 3638.02 &  B8 IV  bl4130Si3856Si5041Si5056 &  B9 III  bl4130Si3856Si5041Si5056 &  & 458330395554092160\\
15303061 & 31.11 & 58.59 & 0.05 & -0.02 & 0.59 & 3633.01 &  B9 II-III  bl4130Si3856Si5041Si5056Cr4172 &  B9.5 III  (bl4130)Si3856Si5041Si5056 &  & 506958083995671424\\
15303088 & 31.09 & 58.35 & 1.54 & 0.12 & 0.58 & 1878.73 &  A0 II-III  bl4130Si3856Eu4205Cr4172 &  B9.5 III-IV  (bl4077)bl4130Si3856Si5056Eu4205Cr4172 &  & 506942312875851264\\
15304172 & 32.93 & 58.53 & 1.54 & 0.09 & 0.59 & 2048.16 &  B9 IV-V  bl4077bl4130Si3856Si5041Si5056Cr4172 &  B9.5 IV-V  bl4077bl4130Si3856Si5041Si5056Cr4172 &  & 506818927055051776\\
107803140 & 87.98 & 20.49 & 0.70 & -0.09 & 0.56 & 953.54 &  B8 IV  Eu4205Cr4172 &  B8 IV-V  Si5056Eu4205 &  & 3399810865002711424\\
107806193 & 90.39 & 19.58 & 1.17 & -0.00 & 0.57 & 713.63 &  A0 IV-V  Si5056Cr4172 &  A0 IV-V  Si5041Si5056Cr4172 &  & 3374701322257770368\\
107814144 & 86.72 & 21.08 & 0.25 & 0.14 & 0.58 & 1746.30 &  B8 II-III  (bl4077)bl4130 &  B9.5 II-III  (bl4077)bl4130Si5041Si5056 &  & 3400054338109166080\\
107904250 & 89.14 & 20.35 & 1.08 & 0.04 & 0.58 & 1120.49 &  B9 IV  bl4077bl4130Eu4205Cr4172Sr4216 &  B9.5 IV  bl4077bl4130Eu4205Cr4172Sr4216 &  & 3423035715975909888\\
107905154 & 88.38 & 19.62 & -0.26 & 0.18 & 0.58 & 2109.28 &  B8 IV  bl4077bl4130Si3856Cr4172Sr4216 &  B9 IV-V  bl4077bl4130Si3856Si5041Cr4172Sr4216 &  & 3398874703869380736\\
165907022 & 301.37 & 44.12 & 0.17 & -0.02 & 0.58 & 1613.04 &  A1 IV-V  bl4077bl4130Eu4205Cr4172 &  A1 IV-V  bl4077bl4130Eu4205Cr4172 &  & 2081867542833415168\\
166108162 & 66.57 & 33.22 & 0.23 & 0.01 & 0.58 & 1160.96 &  B9 III  bl4077bl4130Si3856Si5041Si5056Eu4205Cr4172 &  B8 IV-V  bl4077bl4130Si3856Si5041Si5056Eu4205Cr4172 & new rotation & 172534338135032576\\
166115176 & 65.94 & 35.13 & 1.80 & 0.25 & 0.58 & 338.80 &  kA2hA3mA6  bl4077bl4130Eu4205Cr4172Sr4216 &  kA2hA5mA8  bl4077bl4130Si5041Eu4205Cr4172Sr4216 &  & 176314665270306944\\
167502103 & 78.34 & 39.00 & 0.77 & -0.05 & 0.58 & 1031.67 &  B8 IV-V  bl4077bl4130Si3856Cr4172 &  B8 IV-V  bl4077bl4130Si3856Cr4172 &  & 188235776501129600\\
204805179 & 93.29 & 34.40 & 0.99 & 0.07 & 0.59 & 1000.69 &  B8 III-IV  bl4077bl4130Si3856Si5041Si5056Cr4172 &  B8 IV-V  (bl4077)bl4130Si3856Si5041Si5056Cr4172 & new rotation & 3452249426397874816\\
204815021 & 94.29 & 36.38 & -0.13 & -0.19 & 0.57 & 1584.86 &  B4 II-III  (bl4130) &  B4 III-IV  Si5041 &  & 3453051764937380096\\
204815135 & 94.81 & 36.39 & 0.02 & -0.07 & 0.59 & 1418.80 & B8 II-III  bl4077bl4130Si5041Si5056Cr4172 &  B8 IV-V  bl4077bl4130Si5041Si5056Cr4172 &  & 3453040838540792960\\
204905041 & 93.96 & 34.16 & 0.47 & 0.05 & 0.57 & 3217.13 & B9 II-III  bl4077bl4130Si5041Si5056 &  B9.5 II-III  bl4077bl4130Si3856Si5041Si5056 &  & 3452198269036345984\\
204908180 & 94.89 & 34.14 & 0.64 & 0.06 & 0.58 & 1071.08 &  A0 IV-V  bl4077(bl4130)Eu4205Cr4172 &  A0 IV  bl4077bl4130Cr4172 &  & 3440503068792793472\\
238012135 & 283.69 & 48.34 & 1.52 & 0.08 & 0.58 & 1653.30 & kA1hA3mA4  bl4077bl4130Si3856Eu4205Cr4172Sr4216 &  A1 V  bl4077bl4130Si3856Eu4205Cr4172Sr4216 & EB & 2131718510982169728\\
240511113 & 260.92 & 15.59 & 0.98 & 0.23 & 0.58 & 635.77 &  kA4hA8mA8  bl4077bl4130Si3856Si5056Sr4216 &  A9 V  bl4077bl4130Si3856Si5056Sr4216 &  & 4543881060489030528\\
240515196 & 260.15 & 14.39 & 0.63 & -0.08 & 0.56 & 1305.56 &  B9.5 IV  Si5056 &  B9.5 IV-V  (bl4130)Si5041Si5056 &  & 4543145182269101312\\
243011020 & 285.39 & 41.87 & 1.60 & 0.09 & 0.58 & 310.99 &  A1 IV-V  bl4077bl4130Si3856Si5041Eu4205Cr4172Sr4216 &  A1 IV-V  bl4077bl4130Si3856Si5041Eu4205Cr4172Sr4216 &  & 2104073451467807232\\
243113025 & 288.63 & 40.45 & 0.94 & 0.04 & 0.59 & 371.96 &  B9 V  bl4077bl4130Si3856Eu4205Cr4172 &  B9 V  bl4077bl4130Si3856Eu4205Cr4172 & EB & 2101314128259814912\\
243113207 & 288.36 & 40.17 & 0.39 & -0.01 & 0.59 & 3937.68 &  B9 V  bl4077bl4130Cr4172 &  B9 IV-V  bl4077bl4130Cr4172 & EB & 2101303373661496704\\
297006249 & 87.21 & 35.04 & 0.66 & -0.08 & 0.58 & 644.47 &  B9 V  bl4077bl4130Si3856Eu4205Cr4172 &  B9 V  bl4077bl4130Si3856Cr4172 &  & 3454789508706967296\\
297007224 & 87.08 & 33.59 & 1.26 & 0.02 & 0.58 & 514.48 &  kA4hA9mA8  bl4077bl4130Si3856Si5041Si5056Eu4205Cr4172Sr4216 &  kA4hA9mF1  bl4077bl4130Si3856Si5041Si5056Eu4205Cr4172Sr4216 & no rotation & 3454347367594873088\\
297007241 & 86.65 & 33.32 & -0.38 & 0.10 & 0.57 & 2402.48 &  B8 II-III  (bl4077)bl4130Si3856Si5056 &  B8 IV  (bl4077)bl4130Si3856Si5056 &  & 3448344068466322048\\
297008011 & 86.43 & 35.05 & 1.39 & -0.03 & 0.59 & 884.34 &  A0 IV  bl4077bl4130Si3856Cr4172Sr4216 &  A0 III-IV  bl4077bl4130Si3856Si5056Cr4172Sr4216 &  & 3455528960339050240\\
  297008020 & 86.39 & 35.05 & 1.51 & 0.08 & 0.59 & 909.82 &  A0 IV-V  bl4077bl4130Si3856Eu4205Cr4172 &  A0 IV  bl4077bl4130Si3856Si5041Cr4172 &  & 3455438113192188544\\
  297010019 & 83.13 & 35.45 & 1.24 & 0.04 & 0.59 & 1955.14 &  B9 II-III  bl4077bl4130Si3856Si5041Si5056Cr4172 &  B8 IV  bl4077bl4130Si3856Si5041Si5056Cr4172 &  & 183081609586291712\\
  297010107 & 83.03 & 35.28 & 0.65 & -0.09 & 0.58 & 1018.56 &  B8 III-IV  bl4077bl4130Si3856Si5041Si5056Cr4172 &  B8 IV-V  bl4077bl4130Si3856Si5041Si5056Cr4172 &  & 183077830015921408\\
  297010223 & 83.79 & 35.10 & 0.80 & 0.03 & 0.58 & 1249.87 &  B8 IV-V  (bl4077)bl4130Si3856Cr4172 &  B9 IV-V  (bl4077)bl4130Si3856Cr4172 &  & 183101911891717376\\
  733507176 & 98.64 & 19.11 & 0.44 & 0.03 & 0.59 & 2362.68 &  B8 II-III  bl4077bl4130Si3856Si5041Si5056 &  B9 III  bl4077bl4130Si3856Si5041Si5056 &  & 3371482669469578752\\
  733508216 & 97.65 & 19.26 & 2.58 & 0.25 & 0.60 & 1318.25 &  A0 III  bl4130Si3856Eu4205Cr4172 &  B9 V  bl4130Si3856Cr4172 &  & 3372241852186145792\\
  733510195 & 94.84 & 19.36 & 0.29 & 0.20 & 0.59 & 2655.07 &  B8 III  (bl4130)Si3856Si5056Cr4172 &  B8 IV  bl4130Si3856Si5041Si5056Cr4172 &  & 3373932454393421824\\
  733511065 & 97.26 & 22.30 & 0.45 & 0.13 & 0.58 & 2624.63 &  B8 III-IV  bl4077bl4130Si3856Si5041Si5056Cr4172 &  B9 III  bl4077bl4130Si3856Si5041Si5056 &  & 3376444421848798976\\
  733514197 & 94.86 & 20.69 & 0.58 & -0.01 & 0.59 & 1240.20 &  B8 III-IV  bl4077Si3856Si5041Si5056Cr4172 &  B8 V  bl4077Si3856Si5056Cr4172 &  & 3375810901289300608\\
  733515133 & 97.14 & 21.58 & 0.23 & 0.00 & 0.58 & 1134.72 &  B9 III-IV  (bl4130)Eu4205Cr4172 &  B9 IV  (bl4130)Cr4172 &  & 3376120482532172288
\enddata
\end{deluxetable*}
\end{longrotatetable}

\bibliography{15ver_bpap.bib}
\end{document}